\documentclass[a4paper,12pt]{article}
\pdfoutput = 1
\usepackage[table]{xcolor}				
\usepackage{colortbl}
\usepackage{booktabs}					
\usepackage{multirow,bigdelim}			
\usepackage{amsmath,amssymb,mathtools}	
\usepackage{mathrsfs}					
\usepackage{amsthm}
\usepackage{bbm,dsfont}					
\usepackage{graphicx}					
\usepackage{subfig}						
\usepackage{float}
\usepackage[pdftex,pdfborder={0 0 0}]{hyperref}	
\usepackage[numbers,sort&compress]{natbib}
\usepackage{esint}

\def\hx{\hat{x}}
\def\hp{\hat{p}}

\def\figsize{0.42\linewidth}
\newcommand{\BenderWu}{\texttt{BenderWu}}
\newcommand{\BWDifference}{\texttt{BWDifference}}
\newcommand{\Mathematica}{\texttt{Mathematica}}
\newcommand{\BWProcess}{\texttt{BWProcess}}
\newcommand{\be}{\begin{equation}}
\newcommand{\ee}{\end{equation}}
\newcommand{\normord}[1]{\langle\langle #1\rangle\rangle}
\newcommand{\braket}[3]{\left\langle #1\left| #2\right|#3 \right\rangle}
\newcommand{\brkt}[2]{\left\langle #1|#2\right\rangle}

\newlength{\xtrawidth}
\newlength{\xtraheight}
\def\mc{\mathcal}

\def\bC{\mathbb{C}}
\def\bF{\mathbb{F}}

\def\bN{\mathbb{N}}

\def\bR{\mathbb{R}}
\def\bZ{\mathbb{Z}}

\def\cB{\mathcal{B}}
\def\cC{\mathcal{C}}
\def\cD{\mathcal{D}}

\def\cH{\mathcal{H}}

\def\cN{\mathcal{N}}
\def\cO{\mathcal{O}}

\def\cS{\mathcal{S}}

\def\sp{\mathsf{p}}

\def\sx{\mathsf{x}}

\numberwithin{equation}{section}		
\numberwithin{table}{section}
\numberwithin{figure}{section}

\graphicspath{{figs/}}

\def\({\left(}
\def\){\right)}

\newcommand{\re}{{\rm e}}
\newcommand{\ri}{{\mathsf{i}}}
\newcommand{\rd}{{\rm d}}

\begin{document}
	
	
\begin{titlepage}
	{}~ \hfill\vbox{ \hbox{} }\break

	\rightline{LPTENS 17/29}
	
	\vskip 1 cm

	\begin{center}
		\Large \bf  High order perturbation theory for difference equations and Borel summability of quantum mirror curves
	\end{center}
	
	\vskip 0.8 cm
	
	\centerline{ Jie Gu$^\dagger$, Tin Sulejmanpasic$^\star$}
	
	\vskip 0.2in
%
	\begin{center}{\footnotesize
			\begin{tabular}{c}
			{\em $^\dagger$Laboratoire de Physique Th\'eorique \& $^\star$Institut de Physique Th\'eorique Philippe Meyer,}\\
			{\em \'Ecole Normale Sup\'erieure \& PSL Research University} \\
			{\em 24 rue Lhomond, 75231 Paris Cedex 05, France.} \\[2ex]
			\end{tabular}
		}\end{center}

		\setcounter{footnote}{0}
		\renewcommand{\thefootnote}{\arabic{footnote}}
		\vskip 60pt
		\begin{abstract}
			We adapt the Bender-Wu algorithm to solve perturbatively but very efficiently the eigenvalue problem of ``relativistic'' quantum mechanical problems whose Hamiltonians are difference operators of the exponential-polynomial type. We implement the algorithm in the function \BWDifference{} in the updated \Mathematica{} package \BenderWu{}. With the help of \BWDifference{}, we survey quantum mirror curves of toric fano Calabi-Yau threefolds, and find strong evidence that not only are the perturbative eigenenergies of the associated 1d quantum mechanical problems Borel summable, but also that the Borel sums are exact.
		\end{abstract}
		
		{
			\let\thefootnote\relax
			\footnotetext{$^\dagger$ jie.gu@lpt.ens.fr, $^\star$ tin.sulejmanpasic@gmail.com}
		}
		
	\end{titlepage}
	\vfill \eject

	\newpage
	\tableofcontents
	
	\section{Introduction}
	
	It is usually very rare to have an exact solution to a quantum mechanical problem. Most quantum mechanical systems are either solved numerically or using some approximation scheme, typically relying on some small parameter. The most famous and general approximation scheme is the perturbative expansion around the Planck constant $\hbar$. Perhaps surprisingly however, the generic expansion coefficients grow factorially with the order, rendering the series badly divergent, which calls into question the meaning of the perturbative expansion itself. Enter the resurgence theory of \'{E}calle, an idea that a proper definition of the complete solution requires the inclusion of terms non-perturbative in the coupling which, upon proper definition, are believed to cure all ambiguities and pathologies associated with the pathological series expansion. See for example \cite{costin2008asymptotics,Marino:2012zq}, and more comprehensive references in \cite{marino2015:instantons}.
	
	Early connection of this interplay were noticed independently by Zinn-Justin and Bogomolny, when considering the contributions of instanton--anti-instanton pair to the partition function \cite{Bogomolny:1980ur,ZinnJustin:1981dx}. They proposed that such a pair is ill-defined itself, and upon a certain  ---somewhat ad hoc--- prescription (the Bogomolny--Zinn-Justin or BZJ prescription in the literature), contains an ambiguity of the same kind that exists in the Borel summation of the perturbation theory. They showed that indeed this ambiguity between perturbative and non-perturbative contributions cancel to leading order. Recently however, the ad-hoc BZJ prescription found an explanation in terms of Lefshetz thimble decomposition \cite{Behtash:2015kna,Behtash:2015kva,Behtash:2015zha,Kozcaz:2016wvy,Dunne:2016jsr,Fujimori:2017osz}. Furthermore these ideas led to methods for solving the Schr\"odinger equation, such as uniform WKB \cite{alvarez2000-cubic,alvarez2000-generic,alvarez2002-quartic,alvarez2004-doublewell,Dunne:2014bca}, exact WKB \cite{voros1983return,aoki1991bender,dillinger1993resurgence,kawai2005algebraic}. We also mention a fresh perspective on the problem of Borel summation \cite{Serone:2016qog,Serone:2017nmd} in which it was shown that in quantum mechanics perturbation theory can be recast in a form which completely captures nonperturbative physics.
	
	On the other hand, resurgence in quantum field theory was discouraged due to the discovery of another source of factorial growth of the perturbation series: the 't Hooft renormalons \cite{tHooft:1977xjm}, which occurs because of the running of the coupling, and has no analogue in the quantum mechanical systems and ordinary differential Sch\"odinger equation. Furthermore, the ambiguities coming from the renormalons did not seem to be a result of semiclassical configurations such as instantons. This stymied works in this direction for a long time, and it became widely believed that resurgence is not operative in QFTs on general grounds.
	
	This changed recently due to two parallel but distinct ideologies. On the one hand, \"{U}nsal and Argyres \cite{Argyres:2012vv,Argyres:2012ka} conjectured that renormalon singularities have a semi-classical explanation if the problem is approached from the regime of weakly coupled theory via the idea of adiabatic continuity \cite{Unsal:2007jx,Unsal:2008ch,Shifman:2008ja}. Indeed in such regimes it was shown that renormalon singularities disappear \cite{Anber:2014sda}, and resurgence is likely operative. However this is difficult to test as no access to high orders of perturbation theory is typically available in QFTs. Nevertheless certain 1+1D models, when dimensionally reduced to quantum mechanics via the special kind of compactification, has weak-strong coupling adiabaticity and resurgent structure \cite{Dunne:2012ae,Dunne:2012zk,Cherman:2013yfa}. Resurgence is likewise useful in quantum field theories without renormalon singularities, for instance the Chern-Simons theory \cite{Gukov:2016njj,Marino:2012zq} and certain supersymmetric field theories. Relatedly resurgence also finds its use in topological string theories, where Borel resummation and resurgence techniques have been used to explore non-perturbative contributions and to turn the asymptotic series of topological string free energy into a finite function \cite{Marino:2006hs,Marino:2007te,Marino:2008ya,Marino:2008vx,Pasquetti:2009jg,Klemm:2010tm,Drukker:2011zy,Garoufalidis:2010ya,Aniceto:2011nu,Schiappa:2013opa}, culminating in  \cite{Santamaria:2013rua,Couso-Santamaria:2014iia,Couso-Santamaria:2015hva,Couso-Santamaria:2016vcc,Couso-Santamaria:2016vwq}.
	
	
	Since resurgence is tightly connected with high orders of perturbation theory, it is of immense practical use to have an efficient way to computer high orders of perturbation theory. Recently in \cite{Sulejmanpasic:2016fwr} a \texttt{Mathematica} package called \texttt{BenderWu} was developed using the method originally used by C.~M.~Bender and T.~T.~Wu \cite{Bender:1990pd} for an anharmonic oscillator, which efficiently computes symbolic perturbative solutions to a generic one dimensional quantum mechanical problem with the Hamiltonian of the form
	\begin{equation}\label{eq:H-differential}
		\cH = -\frac{\hbar^2}{2m}\frac{\partial^2}{\partial x^2} + V(x) \ ,
	\end{equation}
	a second order differential operator, where $V(x)$ is an arbitrary non-singular potential, around one of its harmonic minima. 

	Many quantum mechanical problems also exist whose Hamiltonians are difference operators. They can be regarded as the relativistic version of ordinary quantum mechanical systems, for instance, the relativistic Toda lattices \cite{MR1090424}, the elliptic Ruijnaars-Schneider systems \cite{MR851627,MR887995}, the cluster integral systems \cite{Goncharov2011}, and etc. 
	A particular type of relativistic quantum mechanical systems that has recently attracted a lot of attention is quantum mirror curves, and their studies have been extremely fruitful. Consider topological string theory whose target space is a toric Calabi-Yau threefold. The mirror curve to the threefold is the moduli space of the branes compatible with the toric structure \cite{Aganagic:2001nx}. The quantisation of the mirror curve gives rise to Hamiltonian operators of the type
	\be\label{eq:Hamiltonian}
		\cH(\sx,\sp)=\sum_{(r,s)\in I}a_{r,s}e^{r\sx+s\sp}\ ,  \quad a_{r,s} \in \bR \ ,
	\ee
	where $I$ is a finite set of integer pairs, and $\sx, \sp$ satisfy the canonical commutation relation $[\sx,\sp] = \ri \hbar$. The wave-functions to these Hamiltonians are related to the open topological string partition function associated to the branes \cite{Aganagic:2003qj}\footnote{More general branes and the quantisation of their moduli space can also be considered \cite{Aganagic:2012jb}.}. It is later understood that the quantum mirror curve is more closely related to the refined topological string in the Nekrasov-Shatashvili limit \cite{Nekrasov2009a}. The quantum mirror curve defines a spectral problem, whose quantum-corrected WKB periods coincide with the quantum deformation of the periods of the Calabi-Yau, while the latter determine the NS topological string free energy $F^{\rm NS}$ via the so-called quantum special geometry relation \cite{Mironov:2009uv,Mironov:2009dv,Aganagic:2011mi}.
	
	The exact solution to the spectral problem, however, remained elusive until \cite{Grassi:2014zfa}. Naively one would conjecture that the spectral problem is solved by the Sommerfeld-type quantisation condition
	\begin{equation}\label{eq:4d-quantisation}
		\frac{\partial F^{\rm NS}(\vec{a},\hbar)}{\partial a_i} = 2\pi (k_i + 1/2) \ ,\quad k_i \in \bZ_{\geq 0} \ ,
	\end{equation}
	where $k_i$ are the levels of the eigenenergies, and $\vec{a} = (a_i)$ are the quantum periods. The equation \eqref{eq:4d-quantisation}, nevertheless, cannot be the full story, as the l.h.s., which can be understood as the quantum phase space, have poles whenever $\hbar$ is $2\pi$ multiplied by a rational number. Important non-perturbative corrections were first found in \cite{Kallen:2013qla} to cancel the poles, which, after the numerical work \cite{Huang:2014eha} that reveals more subtle corrections are needed, led to the exact spectral theory for quantum mirror curves \cite{Grassi:2014zfa,Codesido:2015dia}, followed by a detailed study of wave-functions \cite{Marino:2016rsq,Marino:2017gyg}, especially in the special case when $\hbar = 2\pi$ (see related works \cite{Kashani-Poor:2016edc,Sciarappa:2017hds,Kashaev:2017zmv}). One amazing feature of the spectral theory is that it also defines conjecturally a non-perturbative completion of topological string free energy in the conifold frame, which coincides with the results of resurgence analysis \cite{Couso-Santamaria:2016vwq}. This conjecture was proved in a special example in certain limit in \cite{Bonelli:2016idi}. See review \cite{Marino:2015nla} and related works \cite{Kashaev:2015kha,Laptev:2015loa,Marino:2015ixa,Kashaev:2015wia,Gu:2015pda,Codesido:2016ixn,Bonelli:2017ptp}. Furthermore, it has recently become clear that the quantum mirror curve is the quantum Baxter equation of the cluster integrable system \cite{Goncharov2011} associated to the toric Calabi-Yau threefold. Inspired by an elegant reformulation \cite{Wang:2015wdy} of the quantisation condition in \cite{Grassi:2014zfa}, a conjectural exact quantisation condition for the cluster integrable system is also written down \cite{Franco:2015rnr,Hatsuda2015}. The interplay between the quantisation conditions for quantum mirror curve and those for cluster integrable system led to an interesting set of relations for BPS invariants of the Calabi-Yau \cite{Sun:2016obh}, and they were proved in a special category of examples in \cite{Grassi:2016nnt}.
	
	To study these systems, we will generalise the algorithm presented in \cite{Sulejmanpasic:2016fwr} to difference equations of type \eqref{eq:Hamiltonian} and study their spectrum. We added to the mathematica package \texttt{BenderWu}\footnote{The most up-to-date \BenderWu{} package is available at:\\ \color{blue}\hyperref[http://library.wolfram.com/infocenter/MathSource/9479/]{http://library.wolfram.com/infocenter/MathSource/9479/}} of \cite{Sulejmanpasic:2016fwr} a function called \texttt{BWDifference}, which computes efficiently perturbative solutions to one dimensional quantum mechanical problems whose Hamiltonian is a difference operator of the exponential-polynomial type given in \eqref{eq:Hamiltonian}. This allows us to study the spectral problem of quantum mirror curve perturbativelly to a very high order ($\ge100$) in $\hbar$. 
	
	When the toric Calabi-Yau threefold is fano, the Hamiltonian operator arising from the quantisation of mirror curve is unique. Y. Hatsuda \cite{Hatsuda:2015fxa} argued that in the case of one particular toric fano Calabi-Yau threefold, the local $\bF_0$, the perturbative eigenenergies of the Hamiltonian operator are Borel summable and that the Borel sums of the perturbative eigenenergies agree well with the numerical spectrum. The study in \cite{Hatsuda:2015fxa} was up to 36 orders in $\hbar$. With the \BWDifference{} function we are able to extend the study of the local $\bF_0$ to $100$ orders in $\hbar$, and confirm that the Bore-Pade partial sums continue to converge to the exact (numerical) result.
	
	Furthermore we study the perturbative solutions to the Hamiltonian operator associated to \emph{all} toric fano Calabi-Yau threefolds using the function \texttt{BWDifference} in the \texttt{BenderWu} package, and find strong evidence that the spectrum of all of them is Borel summable and that the Borel sum gives the correct answer.	

%
%
	
	The paper is organized as follows. In the next section we describe the adapted Bender-Wu algorithm that solves perturbatively the Hamiltonian difference operators, and how to use the \Mathematica{} function that implements the algorithm. In Sec.~\ref{sc:GK}, we explain the Hamiltonian operators arising from the quantisation of mirror curve in topological string theory on a toric Calabi-Yau threefold, especially when the Calabi-Yau is fano, before proceeding to provide evidence that the perturbative eigenenergies of Hamiltonians associated to all toric fano Calabi-Yau threefolds are Borel summable. Finally in Sec.~\ref{sc:cons} we conclude and discuss possible future directions. We relegate to the Appendix the derivation of the adapted Bender-Wu algorithm, as well as the explanation of the technical observation that all the Hamiltonians we have considered have a unique classical minimum.

	\section{The Bender-Wu method for difference equations and the \BWDifference{} package}
	\label{sc:BW}
	
	In this section, we first describe the Bender-Wu algorithm adapted to solve the eigenvalue problems of Hamiltonian difference operators, and then explain how to use the function \BWDifference{} in the \BenderWu{} package which implements the adapted Bender-Wu algorithm.
	
	\subsection{The recursion relations}
	
	Let us start with the Hamiltonian difference operator of the following form
\be\label{eq:H-xp}
\cH(\sx,\sp) =\sum_{r,s}a_{r,s}e^{r\sx+s\sp}\;, \quad a_{r,s} \in \bR \ ,
\ee
where $\sx$ and $\sp$ satisfy the commutation relation $[\sx, \sp] = \ri \hbar$. In the coordinate representation, $\sx$ is the multiplication by $x$ and $\sp=-\ri\hbar \partial_x$. We wish to study the eigenvalue problem of $\cH(\sx,\sp)$
\begin{equation}
	\cH(\sx,\sp) \Psi(x) = E \Psi(x) \ .
\end{equation}
The Hamiltonian operator is an self-adjoint operator over the domain $\cD$ which consists of wave-functions $\Psi(x)$ that are not only themselves $L^2(\bR)$ integrable but that $\re^{\sx}\psi$ and $\re^{\sp}\psi $ are also $L^2(\bR)$ integrable. This constraint can be translated to the condition in the coordinate representation (see for instance \cite{Faddeev:2014hia}) that the wave-function $\Psi(x)$ admits an analytic continuation into the strip
\begin{equation}\label{eq:D-1}
\cS_{-\hbar} = \{ x - \ri y \in \bC: 0\leq y < \hbar\} \ ,
\end{equation}
where it is $L^2(\bR)$ along the $x$-axis for any fixed value of $y$, and that the limit
\begin{equation}\label{eq:D-2}
\Psi(x-\ri \hbar + \ri 0) = \lim_{\epsilon\rightarrow 0^+} \Psi(x - \ri \hbar + \ri \epsilon)
\end{equation}
exists.

To make the analysis \`{a} la Bender-Wu, it is convenient to rescale $\sx = \sqrt{\hbar} \hx, \sp = \sqrt{\hbar} \hp$. This scaling would not change the eigenvalue $E$ nor the eigenfunction $\Psi$, provided that $\hx, \hp$ satisfy the commutation relation $[\hx, \hp] = \ri$. In the coordinate representation, $\hp$ is the differential operator $-\ri \partial_x$.
The Hamiltonian operator now reads
\be
\cH\(\sqrt{\hbar}\,\hx,\sqrt{\hbar}\,\hp\)=\sum_{r,s}a_{r,s}e^{\sqrt{\hbar}\,(r\hx+s\hp)}\;.
\ee
Let us further assume that the Hamiltonian as a function has a local minimum at the origin; in other words, $\cH$ in small $\hbar$ expansion has no linear term in $\hx$ or $\hp$. If this is not the case we can always use a canonical transformation which takes $(x, p)\rightarrow (x+x_0, p+p_0)$ to achieve this, which amounts to the redefinition of $a_{r,s}$\footnote{Note that this canonical transformation also affects the wave-function $\varphi(x)\rightarrow e^{ip_0 x}\varphi(x-x_0)$.}.

Now let us expand the operator in powers of $\hx$ and $\hp$. Up to an overall constant, we get
\be\label{eq:H-hexpansion}
\cH(\sqrt\hbar\, \hx,\sqrt{\hbar}\,\hp)=\sum_{r,s}a_{r,s} +\frac{\hbar}{2}\(A\hx^2+B\hp^2+C(\hx \hp+ \hp \hx)\)+\cO(\hbar^{3/2})\;,
\ee
where
\be\label{eq:ABC}
A=\sum_{r,s}r^2 a_{r,s}\;,\qquad B=\sum_{r,s} s^2 a_{r,s}\;,\qquad C=\sum_{r,s} r s\:a_{r,s}\;.
\ee

The eigenvalue equation for $\cH$ now reads
\be\label{eq:DiffEq_BWForm}
\cH\(\sqrt{\hbar}\,\hx, \sqrt{\hbar}\,\hp\)\Psi(x)=E\Psi(x)\;.
\ee
We wish to solve this equation perturbatively in the expansion of small $\sqrt{\hbar}$. We show in the Appendix that the energy $E$ and the wave-function $\Psi(x)$ have the following expansion
\be\label{eq:Psi_E_expansion}
\Psi(x)=e^{i\frac{x^2}{2}\alpha}\sum_{l=0}^\infty\sum_{k\geq 0} 
\tilde A_l^k \frac{\psi_k(x/\xi)}{\sqrt{k!}}(\hbar/2)^{l/2}\;,\qquad E=\sum_{l=0}^\infty E_{l-2} \hbar^{l/2}\;.
\ee
where
\be\label{eq:alpha-xi}
\alpha=-C B^{-1}\;, \qquad \xi=\left(\frac{B^2}{AB-C^2}\right)^{1/4}\;,
\ee
and where $\psi_k(x)$ is the level $k$ normalized wave-function of a harmonic oscillator with unit mass and frequency. The prefactor $\re^{\ri \alpha x^2/2}$ of wave-function expansion comes from another canonical transformation that makes the second term in the small $\hbar$ expansion of $\cH(\sqrt{\hbar}\,\hx, \sqrt{\hbar}\,\hp)$ \eqref{eq:H-hexpansion} into the Hamiltonian of a harmonic oscillator.

In the Appendix we give the detailed derivation of an algorithm that solves recursively the expansion coefficients $E_{l-2}, \tilde{A}_l^k$. To summarise, we find that  in the lowest orders,
\begin{equation}
	E_{-2} = \sum_{r,s} a_{r,s}
\end{equation}
is the classical energy, $E_{-1} = 0$, and 
\begin{equation}
	E_0 = 2\nu+1 \ , \quad \nu \in \bN_0 \ ,
\end{equation}
where the non-negative integer $\nu$ specifies the level of the eigenenergy. Fixing the level $\nu$, one finds in the lowest orders for the wave-function
\begin{equation}\label{eq:A-1}
	\tilde{A}^\nu_0 = 1 \ , \quad \text{and}\quad \tilde{A}^k_0 = 0 \ ,\quad k \neq \nu \ ,
\end{equation}
where setting $\tilde A^\nu_0$ to unity is a normalization choice. Furthermore, we can normalize the wave-function so that
\begin{equation}\label{eq:A-2}
	\tilde{A}^\nu_l = 0 \ ,\quad l\geq 1 \ .
\end{equation}
To obtain higher order solutions, we first define
\be
\epsilon_l=\frac{2^{l/2}E_l}{\sqrt{AB-C^2}}\ ,\quad \tilde{a}_{r,s} = \frac{a_{r,s}}{\sqrt{AB-C^2}}\ ,
\ee
and
\be
	\beta(r,s) = (r-CB^{-1} s)\xi + \ri s \xi^{-1} \ .
\ee
Then assuming all the coefficients $\tilde{A}^k_{l'}$ and $\epsilon_{l'}$ are known for $l' < l$, the coefficients $\tilde{A}^k_l$ and $\epsilon_l$ can be computed from the following recursive relations respectively,
\begin{multline}\label{eq:rec_tildeA}
\tilde A_{l}^{k}=\frac{1}{2(k-\nu)}\Bigg(-\sum_{q=2}^{\lfloor\frac{l+2}{2}\rfloor}\sum_{r,s}\tilde a_{r,s} \frac{|\beta|^{2q}}{q!}\frac{1}{2^{q}}F(-k,-q;1;2)\tilde A_{l+2-2q}^{k}
\\-\sum_{q=0}^{\lfloor\frac{l+2}{2} \rfloor}\sum_{3\le n+2q\le l+2}\sum_{r,s}\tilde a_{r,s} \frac{|\beta|^{2q}}{n! q!}\frac{1}{2^{q}}\Bigg(\beta^n{\frac{k!}{(k-n)!}}F(-k+n,-q;1+n;2)\tilde A_{l+2-n-2q}^{k-n}\\+\bar\beta^nF(-k,-q;1+n;2)\tilde A_{l+2-n-2q}^{k+n}\Bigg)+\sum_{n=1}^{l-1}\epsilon_{n}\tilde A_{l-n}^k\Bigg)\;, k\neq \nu \ .
\end{multline}
\begin{multline}\label{eq:rec_epsilon}
\epsilon_l=\sum_{q=2}^{\lfloor\frac{l+2}{2}\rfloor}\sum_{r,s}\tilde a_{r,s} \frac{|\beta|^{2q}}{q!}\frac{1}{2^{q}}F(-\nu,-q;1;2) \delta_{l+2,2q}
\\+\sum_{q=0}^{\lfloor\frac{l+2}{2} \rfloor}\sum_{3\le n+2q\le l+2}\sum_{r,s}\tilde a_{r,s} \frac{|\beta|^{2q}}{n! q!}\frac{1}{2^{q}}\Bigg(\beta^n \frac{\nu!}{(\nu-n)!}F(-\nu+n,-q;1+n;2)\tilde A_{l+2-n-2q}^{\nu-n}\\+\bar\beta^nF(-\nu,-q;1+n;2)\tilde A_{l+2-n-2q}^{\nu+n}\Bigg)\;.
\end{multline}
From the recursion relation \eqref{eq:rec_tildeA} and the initial condition \eqref{eq:A-1} one also finds that $\tilde{A}^k_l = 0$ whenever $k>3l+\nu$.

%
%

We have in fact programmed a function called \BWDifference{} for \texttt{Mathematica} which computes the expansion coefficients $\tilde{A}^k_l, \epsilon_l$ automatically and added it to the updated \BenderWu{} package \cite{Sulejmanpasic:2016fwr}. Before we proceed to explain how the function can be used, we would like to make three claims here about the structure of the perturbative eigenenergies and wave-functions:
\begin{enumerate}
\item There is a unique perturbative solution (up to the normalization constant) of the form \eqref{eq:Psi_E_expansion} for any given level number.
\item Energy expansion contains only powers of $\hbar$, not powers of $\sqrt{\hbar}$.
\item The perturbative wave-function can always be constructed to obey
\begin{equation}\label{eq:Psi-phase}
	\Psi_\nu(x,\sqrt{\hbar})=(-1)^\nu\Psi_\nu(-x,-\sqrt{\hbar}) \ ,
\end{equation}
to every order in perturbation theory.
\end{enumerate}

To prove \emph{claim (i)}, consider the difference equation of the form \eqref{eq:DiffEq_BWForm}. Let us show that this equation cannot have two solutions with the same eigenvalue, both of which reduce to harmonic oscillator solutions as $\hbar\rightarrow 0$. Indeed if this were the case, the two solutions must be orthogonal to each other. But this would mean that in the $\hbar\rightarrow 0$ limit, the two solutions reduce to orthogonal harmonic oscillator solutions with different eigenenergies. This violates the assumption that they have the same eigenvalue. Hence we conclude that only one such solution exists. We can also see that this is the case from the recursion equations \eqref{eq:rec_tildeA}, \eqref{eq:rec_epsilon}, as choosing the coefficients\footnote{This choice is just a choice of normalization.} $\tilde A_l^\nu$ uniquely fixes the solution.


Now let us go back to \eqref{eq:DiffEq_BWForm} and prove the \emph{claim (ii)}. One easy way to see this is to notice that the eigenenergies of $\cH(\sx,\sp)$ have an expansion in $\hbar$ not in $\sqrt{\hbar}$, while the spectrum of $\cH(\sx,\sp)$ should be identical to that of $\cH(\sqrt{\hbar}\,\hx, \sqrt{\hbar}\,\hp)$. Now if we perform the transformation $\sqrt{\hbar}\rightarrow -\sqrt{\hbar}$ and $x\rightarrow -x$, we find that the function $\tilde\Psi(x,\sqrt{\hbar})=\Psi(-x,-\sqrt{\hbar})$ is a solution of the same difference equation with the eigenenergy $E(-\sqrt{\hbar})$. Then we have that
\begin{align}
&\braket{\tilde\Psi}{\cH}{\Psi}=E(\sqrt{\hbar})\brkt{\tilde\Psi}{\Psi}\\
&\braket{\Psi}{\cH}{\tilde\Psi}=E(-\sqrt{\hbar})\brkt{\Psi}{\tilde\Psi}\;.
\end{align}
By complex conjugating the second equation, and subtracting from the first we get that either $\brkt{\tilde\Psi}{\Psi}=0$ or $E(\sqrt{\hbar})=E(-\sqrt{\hbar})$. However we also know that $\Psi(-x,-\sqrt{\hbar})$ and $\Psi(x,\sqrt{\hbar})$ cannot be orthogonal, because they reduce to the same harmonic-oscillator solution in the $\hbar\rightarrow 0$ limit. Hence we must have $E(\sqrt{\hbar})=E(-\sqrt{\hbar})$, i.e. energy must be an even function in $\sqrt{\hbar}$, which means that the eigenvalue series expansion is in even powers of $\sqrt{\hbar}$ only. 

\emph{Claim (iii)} immediately follows from the above. Since $\Psi_\nu(x,\hbar)$ and $\Psi_\nu(-x,-\hbar)$ are wave-functions of the same eigenenergy of level $\nu$, we can construct a new wave-function of again the same eigenenergy,
\begin{equation}
	\tilde{\Psi}_\nu(x,\hbar) = \Psi_\nu(x,\hbar)+ (-1)^{\nu}\Psi_\nu(-x,-\hbar)
\end{equation}
and it satisfies the condition \eqref{eq:Psi-phase}. This parity condition implies that 
%
\be\label{eq:A-van}
\tilde A_{l}^k=0\;,\quad\text{if} \quad (-1)^{l+k+\nu}=-1\ ,
\ee
which is compatible with the initial condition \eqref{eq:A-1} that we choose. From the point of view of the recursion calculation, if the above condition on $\tilde{A}_l^k$ is satisfied for all $l<\tilde l$, then by virtue of \eqref{eq:rec_tildeA} we have that $\tilde A_{\tilde l}^k$ for $(-1)^{\tilde l+k+\nu}=-1$ is given entirely by coefficients which vanish, and hence they vanish themselves. 

Incidentally, from \eqref{eq:rec_epsilon} we can see that if $l$ is odd, the r.h.s. contains coefficients which all vanish by \eqref{eq:A-van}, confirming the claim that only even powers of $\sqrt{\hbar}$ appear in the expansion of $E$.

\subsection{How to use the \BWDifference{} function}

Here we present the \BWDifference{} function which is incorporated into the updated \BenderWu{} \cite{Sulejmanpasic:2016fwr} package of \Mathematica. This function solves perturbatively the difference equation of the form
\be
\cH(X,P)\Psi_\nu(x)= E_\nu\Psi_\nu(x)
\ee
were $\nu$ is the level number and $H(X,P)$ is the ``Hamiltonian'' which depends on the momentum and coordinate displacement operators $X=e^{\sqrt{\hbar}x}$ and $P=e^{\sqrt{\hbar}p}$ (with $p=-i\partial_x$), in the polynomial manner, i.e. that
\be
H(X,P)=\sum_{r=r_{\text{min}}}^{s_{\text{max}}}\sum_{s=s_{\text{min}}}^{s_{\text{max}}}c_{r,s}\normord{X^rP^s}
\ee
for integer $r$ and $s$ (note that these can be negative as well). The $\normord{\dots}$ indicates an ordering of $X$ and $P$. A conventional ordering which renders the operator $H(X,P)$ Hermitian is given by
\be
\normord{X^rP^s}\equiv e^{\sqrt{\hbar}(rp+sx)}\;.
\ee
This ordering is assumed by the \BWDifference{} function. Furthermore the \BWDifference{} function assumes that at $X=P=1$ (i.e. $x=p=0$) the classical function $H(X,P)$ attains (at least a local) minimum.

The \BWDifference{} function produces a perturbative expansion of ``energy'' $E$ at level $\nu$ and an unnormalized wave-function $\Psi(x)$, of the form given in \eqref{eq:Psi_E_expansion}. As we have shown in the previous section, the energy is always in powers of $\hbar$, not $\sqrt{\hbar}$. This means that all $E_l$ in equation \eqref{eq:Psi_E_expansion} vanish whenever the $l$  is odd. For this reason the code returns only even coefficients of $E$, i.e. returns $E_{2n}$. From now on when we talk about the ``order'' of the perturbative expansion we will mean the number $n$, rather than the order of $\sqrt{\hbar}$, for which we reserve the letter $l$. Now note that $n=-1$ is the leading order (i.e. classical energy) which is identical to $E_{-2}=\cH(X=1,P=1)$ and is of order $1/\hbar$ in our convention.

In order to access the \BWDifference{} function, one must first install the \BenderWu{} package bundled with this work. Alternatively the most up-to-date version can be downloaded at 

\begin{center}
{\color{blue}\hyperref[http://library.wolfram.com/infocenter/MathSource/9479/]{http://library.wolfram.com/infocenter/MathSource/9479/}}
\end{center}

After following the installation instructions, the package must be loaded via the command
\begin{verbatim}
<<"BenderWu`"
\end{verbatim}
This allows the user to access all the functions in the \BenderWu{} package, in particular the \BWDifference{} function relevant for this work. 

Now let us see how the \BWDifference{} function works. It takes in four essential arguments: the form of the Hamiltonian $\cH(X,P)$, the name of the two variables $X,P$ as a list of two elements, i.e.~\texttt{\{X,P\}}, the level $\nu$, and the order $l_{\text{max}}$ to which the energy $E_\nu$ shall be computed. The typical syntax is given by 

\begin{verbatim}
BWDifference[X+P+1/(XP),{X,P},2,5]
\end{verbatim}
which computes the perturbative expansion of the second level, to the 5$^\text{th}$ order in $\hbar$.

Once the computation is done, the function returns a list with three elements. The first element is the list of coefficients $\{E_{-2},E_0,E_2, E_4,E_6, E_{8},E_{10}\}$, while the second is a matrix of coefficients $A_l^k$ where the $l$-index denotes the rows and the $k$-index the columns. The third element is not important for the user, and only serves for proper functioning of the function \BWProcess{}, which was introduced in \cite{Sulejmanpasic:2016fwr}. Hence if we execute the command 
\begin{verbatim}
BWDifference[X+P+1/(XP),{X,P},5][[1]]
\end{verbatim}
we will get a list of perturbation series coefficients $\epsilon_l$, which in this case is
\be
\left\{3,\frac{5 \sqrt{3}}{2},\frac{77}{72},\frac{145}{432 \sqrt{3}},-\frac{3077}{279936}\right\}
\ee
Alternatively one can use an option \texttt{Output->"Energy"} instead, i.e.
\begin{verbatim}
BWDifference[X+P+1/(XP),{X,P},2,5,Output->Energy]
\end{verbatim}
with the same outcome as before. 

However a better way to use the code is to assign the output to a variable, and use the function \texttt{BWProcess} introduced already in the original \BenderWu{} package \cite{Sulejmanpasic:2016fwr} to control the output without having to recompute the expansion. In other words the benefit of using the \BWProcess{} function is that one can make a computation to a high order once, and use the \BWProcess to analyze the result without having to recompute the expansion. For example, if we call the line
\begin{verbatim}
BW=BWDifference[X+P+1/(XP),{X,P},2,20];
\end{verbatim}
it assigns the output of \BWDifference{} to a variable \texttt{BW}, and hence contains all the perturbative information to order $20$ in $\hbar$. In order to output the energy coefficients, we can simply call
\begin{verbatim}
BWProcess[BW]
\end{verbatim}
which produces the output
\[
\left\{3,\frac{5 \sqrt{3}}{2},\frac{77}{72},\frac{145}{432 \sqrt{3}},-\frac{3077}{279936},\dots\right\}
\]
where the dots stand for the terms not written. Often the computation will involve many terms, and the output can be quite bulky. Therefore the \BWProcess{} function has an option which allows the user to display only the limited order, for example
\begin{verbatim}
BWProcess[BW,Order->5]
\end{verbatim}
Furthermore the \BWProcess{} function can be used to specify the lower and upper bounds of the perturbative order, as in
\begin{verbatim}
BWProcess[BW,Order->{5,10}]
\end{verbatim}
which gives an output
\[
\left\{\frac{5 \sqrt{3}}{2},\frac{77}{72},\frac{145}{432 \sqrt{3}},-\frac{3077}{279936},\frac{25745}{1679616
   \sqrt{3}},-\frac{1621121}{906992640}\right\}
\]

To obtain the wave-function coefficients, all we need to do is to use the option \texttt{Output->"WaveFunction"}. For instance, calling the line
\begin{verbatim}
BWProcess[BW,Output->"WaveFunction", Order->5]//MatrixForm
\end{verbatim}
produces an output
\[
\left(
\begin{array}{ccccccccccccccc}
 0 & 0 & 1 & 0 & 0 & 0 & 0 & 0 & 0 & 0 & 0 & 0 & 0 & 0 & 0 \\
 0 & 0 & 0 & 0 & 0 & -\frac{10}{3} & 0 & 0 & 0 & 0 & 0 & 0 & 0 & 0 & 0 \\
 0 & 0 & 0 & 0 & 0 & 0 & 0 & 0 & \frac{280}{9} & 0 & 0 & 0 & 0 & 0 & 0 \\
 0 & 0 & 0 & 0 & 0 & \frac{20}{9} & 0 & 0 & 0 & 0 & 0 & -\frac{15400}{27} & 0 & 0 & 0 \\
 0 & 0 & 0 & 0 & 0 & 0 & 0 & 0 & -\frac{392}{9} & 0 & 0 & 0 & 0 & 0 & \frac{1401400}{81} \\
\end{array}
\right)
\]
The first row is $A_0^k$, which is zero except for $k=\nu=2$. The second row is $A_1^k$, the third row $A_1^k$, etc. To get the element $A_3^5$, all we need is take the $(4,6)$ element of this output, i.e.\ calling
\begin{verbatim}
BWProcess[BW, Output -> "WaveFunction"][[4, 6]]
\end{verbatim}
which returns
\[
\frac{20}{9}\;.
\]
We can also use an option \texttt{OutputStyle->"Series"} to output the series \eqref{eq:Psi_E_expansion} for the wave-function. For example writing
\begin{verbatim}
BWProcess[BW, Output -> "WaveFunction",OutputStyle->"Series",Order->1]
\end{verbatim}
produces the following output
\[
1-\frac{1}{18} g x \left(\sqrt{3} x^2-3\right)+\frac{1}{648} g^2 \left(3 x^6-15 \sqrt{3} x^4+45 x^2-5 \sqrt{3}\right)\;,
\]
where $g$ is\footnote{Note that this can also be changed by calling the option \texttt{Coupling->Sqrt[hbar]}.} $\sqrt{\hbar}$, so $g^3$ is of order $\hbar^3$. Note that the prefactor of \eqref{eq:Psi_E_expansion} is not included. To include it use the option \texttt{Prefactor->True}
\begin{verbatim}
BWProcess[BW, Output -> "WaveFunction",OutputStyle->"Series",
         Order->6,Prefactor->True]
\end{verbatim}
	\[
	e^{-\frac{1}{4} \sqrt{3} x^2-\frac{i x^2}{4}} \left(1-\frac{1}{18} g x \left(\sqrt{3} x^2-3\right)+\frac{1}{648} g^2 \left(3 x^6-15 \sqrt{3} x^4+45 x^2-5 \sqrt{3}\right)\right)
	\]
Let us define the wave-function and energy to the 10th order of $\hbar$ with the commands
\begin{verbatim}
psi[x_]:=Evaluate[BWProcess[BW, Output -> "WaveFunction", 
		 OutputStyle -> "Series", Order -> 10, Prefacto r-> "True"]];
epsilon = BWProcess[BW, Output -> "Energy", OutputStyle -> "Series"];
\end{verbatim}
The difference equation for the difference operator $\cH=X+P+\langle\langle1/(XP)\rangle\rangle$, with $X=e^{igx}, P=e^{igp}$, explicitly reads
\be
\psi(x - \ri g)+e^{gx}\psi(x)+e^{-gx- \ri\frac{g^2}{2}}\psi(x + \ri g)=g^2\epsilon \psi(x)\;.
\ee
To verify the above equation to order 20 in $g=\sqrt{\hbar}$, we use execute
\begin{verbatim}
Simplify[Series[psi[x - I g] + Exp[x g] psi[x] + 
    Exp[-x g - I g^2/2] psi[x + I g] - g^2 epsilon psi[x], {g, 0, 20}]]
\end{verbatim}
which returns $o[g^{21}]$, so that the equation is satisfied at least to the 20th order in $g=\sqrt{\hbar}$.

Finally we discuss briefly the option \texttt{Imaginary}. The solution of the difference equation $\psi(x)$ need not be real (up to a constant phase), and the coefficients $\tilde A_l^k$ can have imaginary parts. The example we studied so far returns purely real coefficients $\tilde A_l^k$ (see Appendix \ref{sc:BW-rec}). When the coefficients are not real, the algorithm may slow down significantly, especially if large orders need to be computed. In order to improve this, a refined algorithm is built into the \BWDifference{} function which speeds up the computation when the coefficients are complex by separating the real and the imaginary parts of the coefficients. To switch to the refined algorithm, one needs only to add \texttt{Imaginary->True} in the option list of the \BWDifference{} function. For concrete examples, see the example notebook included in the \BenderWu{} package.

	\section{Application: quantum mirror curves}
	\label{sc:GK}
	
	We describe here the Hamiltonian operators arising from the quantisation of mirror curves in topological string theory on toric fano Calabi-Yau threefolds, and then apply our Bender-Wu algorithm to solve perturbatively the eigenvalue problem of the Hamiltonians.
	
	\subsection{Quantum mirror curves}
	\label{sc:generality}

	Consider topological string theory on a toric Calabi-Yau threefold \cite{Hosono:1994av,Batyrev:1994hm,Hori:2000kt,Cox:2000vi}.
	A toric Calabi-Yau threefold $X_\Sigma$ can be succinctly described by its toric fan $\Sigma$. The toric fan consists of $n_{\Sigma}+3$ 1-cones and the triangulation of the convex hull of the 1-cones. The 1-cones are subject to $n_\Sigma$ linear relations
	\begin{equation}
		\sum_{\alpha=1}^{n_\Sigma+3} \ell_\alpha^{(i)} \bar{v}_\alpha = 0 \ , \quad \quad \ell^{(i)}_\alpha\in \bZ \ , i=1,\ldots, n_{\Sigma} \ .
	\end{equation}
	The Calabi-Yau condition demands that one can always rotate the toric fan so that the endpoints of the 1-cones have coordinates
	\begin{equation}
		\bar{v}_\alpha = (1, r_\alpha, s_\alpha) \ ,\quad\quad  r_\alpha, s_\alpha \in \bZ \ .
	\end{equation}
	It is therefore enough to present the toric fan by the image of the projection onto the plane $(1,\bullet,\bullet)$, a triangulated convex integral polygon whose vertices are
	\begin{equation}
		v_\alpha = (r_\alpha, s_\alpha) \ ,\quad \alpha = 1, \ldots, n_{\Sigma}+3 \ .
	\end{equation}
	We call this image the support of toric fan or simply the fan support, denoted by $N_\Sigma$. A toric Calabi-Yau threefold can have different fan supports which are related to each other by $SL(2,\bZ)\ltimes\bZ^2$
	\begin{equation}\label{eq:isometry}
	\(\begin{array}{c} r_\alpha \\ s_\alpha \end{array}\)
	\mapsto
	\(\begin{array}{c}
		a r_\alpha + b s_\alpha + c_1 \\
		c r_\alpha + d s_\alpha + c_2
	\end{array}\) \ , \quad\quad\begin{pmatrix}
		a & b \\ c & d
	\end{pmatrix} \in SL(2,\bZ) \ , c_{1,2}\in \bZ \ ,
	\end{equation}
	which preserves the linear relation vectors
	\begin{equation}
		\ell^{(i)} = (\ell^{(i)}_\alpha) \ .
	\end{equation}
		
	Mirror symmetry dictates that the free energies of topological string theory on the Calabi-Yau threefold $X_\Sigma$ can be computed from the mirror curve $\cC_\Sigma$, a noncompact Riemann surface, whose Newton polygon coincides with the fan support of $X_\Sigma$. Therefore given the fan support $N_\Sigma$ with vertices $v_\alpha$, the equation of $\cC_\Sigma$ reads
	\begin{equation}\label{eq:mcurve}
		\sum_{\alpha=1}^{n_{\Sigma}+3} a_\alpha \re^{r_\alpha x+s_\alpha y} = 0 \ ,\quad x,y\in \bC \ .
	\end{equation}
	
	The coefficients $a_\alpha$ in the equation \eqref{eq:mcurve} parametrise the complex structure moduli space of the mirror curve. They are not all independent, as three of them can be scaled to one through the $\bC^*$ scalings on $\re^x, \re^y$ and an overall scaling. It is customary to set to 1 three coefficients associated to vertices on the boundary; the number of internal vertices gives the genus $g_\Sigma$ of the mirror curve. Due to physics consideration, the $g_\Sigma$ coefficients associated to internal vertices are called the true moduli, while the remaining coefficients associated to boundary vertices after fixing the $(\bC^*)^3$ scaling are called mass parameters\footnote{With rare exceptions, the topological string on a toric Calabi-Yau threefold engineers a 5d $\cN=1$ supersymetric gauge theory. The true moduli are Coulomb moduli while the mass parameters are either the masses of hypermultiplets or the fugacity of instanton counting.}. 
	
\begin{figure}
		\centering
		\includegraphics[width=0.8\linewidth]{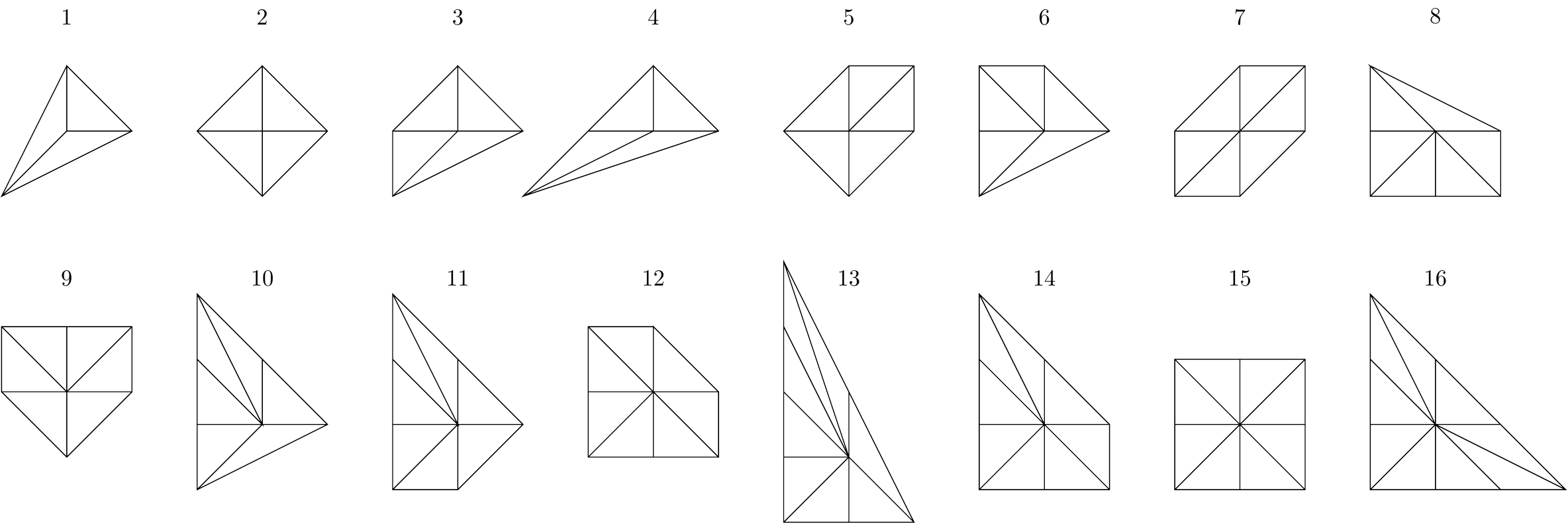}
		\caption{Sixteen reflexive Newton polygons.}\label{fg:reflexive}
	\end{figure}

	In this paper for simplicity we restrict ourselves to fano Calabi-Yau threefolds whose fan supports are reflexive, in other words convex Newton polygons with only one internal vertex. Reflexive 2d polygons have been classified up to the $SL(2,\bZ)$ isometry, and they are listed in Fig.~\ref{fg:reflexive} (see for instance the construction in \cite{batyrev1984higher,koelman1990number}). Since they have a single internal vertex, and it allows for a canonical way of writing down the curve equation by putting the only internal vertex at the origin. For instance, the canonical equation for the first polygon in Fig.~\ref{fg:reflexive} is
	\begin{equation}\label{eq:P2}
		\re^x + \re^y + \re^{-x-y} + u = 0 \ ,
	\end{equation}
	while the second polygon in Fig.~\ref{fg:reflexive} gives
	\be
	\re^x+\re^y+\re^{-x}+\re^{-x-y}+u=0
	\ee
	In these equations $u$ is the true modulus of the model. Note the canonical form still enjoys the $SL(2,\bZ)$ isometry acting on the exponents
	\begin{equation}
		(r_i, s_i) \mapsto (a r_i + b s_i, c r_i + d s_i)\ , \quad \begin{pmatrix}
			a & b \\ c & d
		\end{pmatrix} \in SL(2,\bZ) \ .
	\end{equation}
	
	To quantise the mirror curve, we simply promote the coordinates $x,y$ to quantum operators $\sx,\sp$ satisfying the canonical commutation relation $[\sx, \sp] = \ri\hbar$ through the Weyl quantisation prescription
	\begin{equation}
	\re^{r_i x + s_i y} \mapsto \re^{r_i \sx + s_i \sp} \ .
	\end{equation}
	Here $\hbar$ is assumed to be real. For a genus $g_\Sigma$ mirror curve, one can in principle construct $g_\Sigma$ mutually non-commutative Hamiltonian operators, each associated to a different true modulus \cite{Codesido:2015dia}. The mirror curve of a fano Calabi-Yau threefold is always of genus one, and thus the associated Hamiltonian is unique. It is obtained by taking the l.h.s.\ of the canonical equation of curve, removing the true modulus $u$, and then performing the quantisation procedure. In the example of \eqref{eq:P2}, we get
	\begin{equation}
	\mathcal{H} = \re^{\sx} + \re^{\sp} + \re^{-\sx - \sp} \ .
	\end{equation}
	The $SL(2,\bZ)$ isometry of the Newton polygon then corresponds to canonical transformations on $\sx, \sp$.
	

	In this paper, we are interested in the eigenvalue problem of the Hamiltonian operator associated to a toric fano Calabi-Yau threefold, in the following form\footnote{Whether we consider the perturbative series of the eigenvalue of $\cH$ or its logarithm is a matter of convention. In the results we discuss the expansion of $\mathcal E$. 
	}
	\begin{equation}
		\cH(\sx, \sp) \Psi_\nu(x) = \re^{\mathcal E^{(\nu)}} \Psi_\nu(x) \ ,
	\end{equation}
	where $k$ is the level number. In \cite{Grassi:2014zfa} a conjectural quantisation condition was given using the (refined) topological string free energies to solve exactly the spectrum of $\cH(\sx, \sp)$. In this paper, we are interested in the perturbative solution to the Hamiltonian eigenvalue problem, and we will not need the input of topological string. Clearly the Hamiltonian operator is of the form \eqref{eq:H-xp}, and so its eigenvalue problem can be treated by our \BWDifference{} function. We also call the polynomial of $\re^x,\re^y$ before quantisation the Hamiltonian function $H(x,y)$, and it is the analogue of the classical potential in a nonrelativistic quantum mechanical problem.

	Consider the perturbative expansion of $\mathcal E^{(\nu)}$ in terms of $\hbar$
	\begin{equation}
		\mathcal E^{(\nu)} = \sum_{n=0}^\infty \hbar^n \mathcal E^{(\nu)}_n \ ,
	\end{equation}
	which is an asymptotic series with zero radius of convergence. 
	Hatsuda in \cite{Hatsuda:2015fxa} gave evidence that for the second geometry in the list of Fig.~\ref{fg:reflexive} with the mass parameter set to 1, the Borel sum of the perturbative eigenenergies for finite values of $\hbar$ agrees with the numerical results, implying the Borel summability of the eigenenergy series. We want to expand the exploration in \cite{Hatsuda:2015fxa} to other reflexive geometries with higher precision. The precision of Borel resummation depends crucially on the order of asymptotic series that is included. \cite{Hatsuda:2015fxa} fixed the coefficients of the perturbative eigenenergies by comparing the asymptotic series with numerical eigenenergies computed by numerous small values of $\hbar$, and in this way, \cite{Hatsuda:2015fxa} could only obtain up to order 36 of the perturbative eigenenergies for the said geometry. Our \BWDifference{} function provides a far more efficient way to compute perturbative eigenenergies. For instance, for the same geometry the \BWDifference{} function can easily compute the eigenenergy series at level 0 up to order 100 within 240 seconds on an ordinary desktop computer. This results in an agreement between the Borel sums with the numerical results for $\hbar = \pi$ up to more than 25 digits, compared to only 12 matching digits in \cite{Hatsuda:2015fxa}.
	
	We analysed all sixteen reflexive Newton polygons listed in Fig.~\ref{fg:reflexive}, corresponding to all possible toric fano Calabi-Yau threefolds, for appropriately chosen values of  mass parameters. We find that for each model the poles of the Borel transforms of the perturbative eigenenergies are never located on the positive real axis of the Borel plane, indicating Borel summability. Besides, the Borel sums of the eigenenergies have very good agreement with the numerical results, and the degree of agreement increases consistently when more orders of perturbative series are used in resummation. We therefore confirm and expand to all toric fano Calabi-Yau threefolds the observation in \cite{Hatsuda:2015fxa} that the Borel-Pad\'{e} resummation captures the exact eigen-energies. The details of the results are discussed in the next section.

	\subsection{Results}
	\label{sc:results}
	
	\begin{table}
		\centering\resizebox{\linewidth}{!}{
			\begin{tabular}{>{$}c<{$} >{$}l<{$}}\toprule
				\text{geometry} & \text{Hamiltonian operator} \\\midrule
				F_1 & \cH = \re^{\sx} + \re^{-\sx/2+\sp} + \re^{-\sx/2-\sp} \\
				F_2 & \cH = \re^{\sx} + m_1\re^{-\sx} + \re^{\sp} + \re^{-\sp}  \\
				F_3 & \cH = \re^{\sx} + \re^{-\sx/2+\sp} + \re^{-\sx/2-\sp} + m_1 \re^{-\sx} \\
				F_4 & \cH = \re^{\sx} + \re^{-\sx+\sp} + \re^{-\sx -\sp} + m_1 \re^{-\sx} \\
				F_5 & \cH = \re^{\sx/2 - \sp} + \re^{\sx/2 + \sp} + \re^{-\sx} + m_1 \re^{-\sx/2 + \sp} + m_2 \re^{-\sx/2- \sp} \\
				\rowcolor{gray!20}
				F_6 & \cH = \re^{\sx} + \re^{\sp} + \re^{-\sx - \sp} + m_1 \re^{-\sx} + m_2 \re^{-\sx + \sp} \\
				F_7 & \cH = \re^{\sx/2 - \sp} + \re^{\sx/2 + \sp} + \re^{-\sx} + m_1\re^{\sx} + m_2 \re^{-\sx/2+\sp} + m_3 \re^{-\sx/2-\sp} \\
				\rowcolor{gray!20}
				F_8 & \cH = \re^{\sx} + \re^{\sp} + \re^{-\sx - \sp} + m_1\re^{\sx+ \sp} + m_2 \re^{-\sx} + m_3 \re^{-\sx + \sp} \\
				F_9 & \cH = \re^{\sx+\sp} + \re^{\sx - \sp} + \re^{-\sx} + m_1 \re^{\sx} + m_2 \re^{-\sp} + m_3 \re^{\sp} \\
				\rowcolor{gray!20}
				F_{10} & \cH = \re^{\sx} + \re^{\sp} + \re^{-\sx - \sp} + m_1 \re^{-\sx} +m_2 \re^{-\sx + \sp} + m_3 \re^{-\sx + 2\sp} \\
				\rowcolor{gray!20}
				F_{11} & \cH = \re^{\sx} + \re^{\sp} + \re^{-\sx-\sp} + m_1 \re^{-\sx} +m_2 \re^{-\sx + \sp} + m_3 \re^{-\sx+2\sp} +m_4 \re^{-\sp} \\
				F_{12} & \cH = \re^{\sx/2 - \sp} + \re^{\sx/2 + \sp} + \re^{-\sx} + m_1\re^{-\sx/2+\sp} + m_2\re^{-\sx/2-\sp} + m_3\re^{2\sp} + m_4 \re^{-2\sp} \\
				F_{13} & \cH = \re^{\sx}+\re^{-\sx-2\sp}+\re^{-\sx+2\sp}+m_1 \re^{\sp} + m_2 \re^{-\sp} + m_3 \re^{-\sx-\sp} + m_4 \re^{-\sx + \sp} + m_5 \re^{-\sx}\\
				F_{14} & \cH = \re^{\sx + \sp/2} + m_1\re^{\sx - \sp/2} + \re^{-\sx-3\sp/2} + \re^{-\sx+3\sp/2} + m_2\re ^{-\sp} + m_3\re ^{\sp} + m_4\re ^{-\sx-\sp/2} + m_5\re ^{-\sx +\sp/2} \\
				F_{15} & \cH = \re^{\sx/2-\sp} + \re^{\sx/2+\sp} + \re^{-\sx} + m_1 \re^{\sx} + m_2 \re^{2\sp} + m_3 \re^{-2\sp} + m_4 \re^{-\sx/2+\sp} + m_5\re^{-\sx/2-\sp}\\
				F_{16} & \cH = \re^{\sx/2-\sp} + \re^{\sx/2+\sp} + \re^{-\sx} + m_1 \re^{-\sx/2+\sp} + m_2 \re^{-\sx/2-\sp} + m_3 \re^{2\sp} + m_4 \re^{-2\sp} + m_5 \re^{\sx/2+3\sp} + m_6 \re^{\sx/2-3\sp}\\
				\bottomrule
			\end{tabular}}
			\caption{Hamiltonian operators associated to the 16 reflexive Newton polygons arranged in such a way that with appropriate values of mass parameters they are $\sp$-parity invariant, except for $F_6, F_8, F_{10}, F_{11}$ which are marked out in gray.}\label{tb:Hamiltonians}
		\end{table}	
	
	We first write down in Tab.~\ref{tb:Hamiltonians} the Hamiltonian operators for each of the 16 reflexive Newton polygons listed in Fig.~\ref{fg:reflexive}. It is beneficial if we can rearrange the Hamiltonian operator so that it is invariant under the reflection $\sp \mapsto -\sp$. We call such an operator $\sp$-parity even. From the point of view of perturbative calculation via the \BWDifference{} function, the wave-functions of a $\sp$-parity odd Hamiltonian operator are complex, and the computation is significantly slowed down compared to the cases where wave-functions are real. This problem can be circumvented by turning on the option \texttt{Imaginary->True} in the \BWDifference{} function, which then separates the real and the imaginary parts of complex wave-functions explicitly to cure the slowdown. From the point of view of numerical calculation, when working in the coordinate representation, the $\sp$ operator is $-\ri \hbar \partial/\partial_x$. As a consequence, if the Hamiltonian operator is $\sp$-parity even, the Hamiltonian matrix with entries $\langle n | \cH | m\rangle$ would be real symmetric instead of complex Hermitian, and thus the matrix diagonalisation would be faster. 
	
	Here is an appropriate place to recall the method of numerical calculation of spectrum (see for example \cite{Huang:2014eha}). We choose the basis of wave-functions in the domain of $\cH$ to consist of the eigenfunctions of the quantum harmonic oscillator with both mass and frequency set to 1, i.e.
	\begin{equation}
		\langle x | n \rangle = \psi_n(x) = \frac{1}{\sqrt{2^n n!} (\pi\hbar)^{1/4}} \re^{-\tfrac{x^2}{2\hbar}} H_n\( \frac{x}{\sqrt{\hbar}}\) \ ,\quad n=0,1,\ldots \ .
	\end{equation}
	Here $H_n(x)$ are Hermite polynomials, and they obey the following orthogonality conditions
	\begin{equation}
		\int_{-\infty}^{\infty} \re^{-x^2}H_{n_1}(x+y) H_{n_2}(x+z) \rd x = 2^{n_2} \sqrt{\pi}n_1! z^{n_2 - n_1} L_{n_1}^{n_2 - n_1}(-2y z) \ , \quad n_1 \leq n_2 \ ,
	\end{equation}
	where $L_n^\alpha(z)$ are Laguerre polynomials. Then for the operator $\re^{r \sx + s \sp}$, we have
	\begin{equation}
		\langle n_1 | \re^{r \sx + s \sp} | n_2 \rangle  = \sqrt{n_1! n_2!}\,\re^{\tfrac{|z|^2}{2}} z^{n_1} \bar{z}^{n_2} \sum_{k=0}^{\min(n_1, n_2)} \frac{1}{k!(n_1 - k)!(n_2 - k)!}\frac{1}{|z|^{2k}} \ ,
	\end{equation}
	where
	\begin{equation}
		z = \sqrt{\hbar/2}(r + \ri s) \ .
	\end{equation}
	Clearly the Hamiltonian matrix $\langle n_1 | \cH | n_2 \rangle$ is real and symmetric if and only if every monomial $\re^{r \sx + s \sp}$ is paired with $\re^{r \sx - s \sp}$, in other words, the Hamiltonian operator is $\sp$-parity even.
	
	Among the 16 reflexive Newton polygons, the Hamiltonians of all but four geometries, namely $F_6, F_8, F_{10}, F_{11}$, can be put via a canonical transformation to a form that is $\sp$-parity even for appropriately chosen values of mass parameters. This is the form of the Hamiltonians presented in Tab.~\ref{tb:Hamiltonians}.
	
	When mass parameters are non-negative
	, the Hamiltonian functions for the operators in Tab.~\ref{tb:Hamiltonians} have a unique minimum, as is shown in the Appendix \ref{sc:uniqueness}\footnote{In the case of $F_2$ the mass parameter has to be positive for the minimum to exist.} for real values of $x,y$, which is taken to be the classical ground state. 
	The uniqueness of the classical ground state also indicates the absence of real instantons, and could be related to the Borel summability of the spectrum that we find here.
	
	From the point of view of perturbative solutions, the \BWDifference{} function expands around a minimum of the Hamiltonian function which it assumes to be $(x,y) = (0,0)$. Therefore when the actual minimum $(x,y) = (x_0,y_0)$ is not at the origin, we have to shift the coordinates $x,y$ by hand
	\begin{equation}
		(x,y) \mapsto (x+x_0, y+ y_0)
	\end{equation}
	before feeding the Hamiltonian function into the \BWDifference{} function. Furthermore, the \BWDifference{} function runs much faster if the Hamiltonian function after the shift of coordinates has no irrational coefficients. We can always achieve this by taking appropriate values of mass parameters.
	
	\begin{table}
		\centering
		\begin{tabular}{ *{2}{>{$}c<{$} >{$}l<{$}}}\toprule
			\text{geometry} & \text{mass parameters} & \text{geometry} & \text{mass parameters}\\\midrule
			F_1 & - & F_9 & (2,1,1)\\
			F_2 & (1) & F_{10} & (5/4,1,1)\\
			F_3 & (14) & F_{11} & (3/4,2,1,1/8)\\
			F_4 & (2) & F_{12} & (7/2,1,7/2,1)\\
			F_5 & (7/2,7/2) & F_{13} & (1,1,1/2,1,1/2)\\
			F_6 & (1,2) & F_{14} & (1,1,1,3,3)\\
			F_7 & (1,1,1) & F_{15} & (1,1,1,1,1)\\
			F_8 & (1/4,2,1) & F_{16} & (9/2,1,1/4,9/2,1,1/4)\\
			\bottomrule
		\end{tabular}
		\caption{Choices of mass parameters. An entry $(c_1,c_2,\ldots)$ means the mass parameters take values $(m_1,m_2,\ldots) = (c_1,c_2,\ldots)$. 
		}\label{tb:models}
	\end{table}		

	As we have seen, in order to most efficiently use the \BWDifference{} function, we would like to choose rational values of mass parameters such that
	\begin{itemize}
		\item the Hamiltonian operator is $\sp$-parity even (not applicable to $F_6, F_8, F_{10}, F_{11}$);
		\item the coordinates $(\re^{x_0}, \re^{y_0})$ of the minimum of the Hamiltonian function are rational numbers.
	\end{itemize}
	We choose one set of mass parameters for each geometry satisfying these conditions, and list them in Tab.~\ref{tb:models} ($F_1$ has no mass parameter).
	
	Let us focus for the moment on the polygon $F_2$, which represents the Calabi-Yau threefold called the canonical bundle over the Hirzebruch surface $\bF_0$ or local $\bF_0$, and we set the mass parameter $m_1=1$, as indicated in Tab.~\ref{tb:models}. As already mentioned in Section~\ref{sc:generality}, we can compute the perturbative series of the ground state energy up to order 100 with relative ease. Now given the asymptotic series $\mathcal E^{(\nu)}(\hbar)$, we can compute the Borel transform
	\begin{equation}
		\cB[\mathcal E^{(\nu)}](\zeta) = \sum_{n=0}^{\infty} \frac{E_n^{(k)}}{n!}\zeta^n \ ,
	\end{equation}
	which is a convergent series. The Borel transform may have poles in the $\zeta$-plane, also known as the Borel plane, and the locations of the poles are the actions of the instantons of the relevant quantum mechanical system. If no pole lies on the positive real axis, we can perform the Laplace transformation on the Borel transform
	\begin{equation}
		\cS[\mathcal E^{(\nu)}](\hbar) = \int_{0}^{\infty} \frac{\re^{-\zeta/\hbar}}{\hbar} \cB[\mathcal E^{(\nu)}](\zeta)  \rd \zeta \ ,
	\end{equation}
	which results in an analytic function $\cS[\mathcal E^{(\nu)}](\hbar)$ that is well-defined for finite values of $\hbar$. The function $\cS[\mathcal E^{(\nu)}](\hbar)$ has the property that its expansion around $\hbar = 0$ coincides with the asymptotic series we start with, which is $\mathcal E^{(\nu)}(\hbar)$ in our case. This procedure of obtaining an analytic function out of an asymptotic series is call \emph{Borel resummation}. It is called the \emph{Borel-Pad\'{e}} resummation if $\cB[\mathcal E^{(\nu)}](\zeta)$ is  replaced by the Pad\'{e} approximant $\mc P[\mathcal E^{(\nu)}](\zeta)$ of the Borel transform of a truncated series.
	
\begin{figure}
		\centering
		\includegraphics[width=0.5\linewidth]{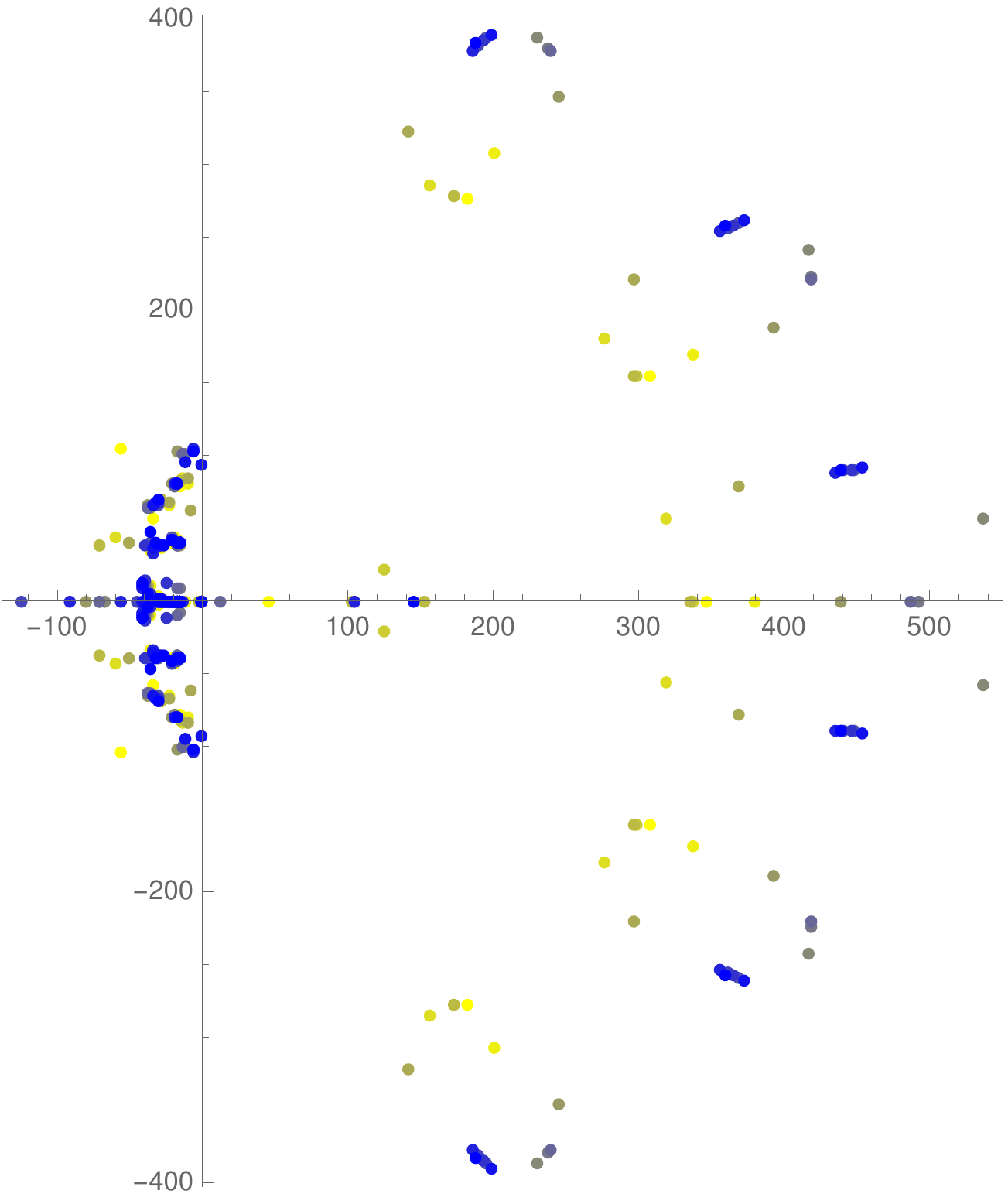}
		\caption{Poles of the Pad\'e approximants $\mc P[\mathcal E^{(\nu)}](\zeta)$  of the Borel transform for the perturbative ground state energy for $F_2$ for orders from 70 to 100. The poles which are more yellow are of lower order, while the poles which are more blue are of higher order Pad\'e approximants.}\label{fg:poles-F2}
	\end{figure}

	To study the Borel plane for the model of local $\bF_0$ with $m_1 = 1$, we plot in Fig.~\ref{fg:poles-F2} the poles of $\mc P[\mathcal E^{(\nu)}](\zeta)$ for the series $\mathcal E^{(\nu)}(\hbar)$ truncated at various orders, from order 70 up to order 100, with poles of lower order series more yellowish while poles of higher order series more blueish. No stable poles of $\mc P[\mathcal E^{(\nu)}](\zeta)$ accumulate along the positive real axis, in accord with the observation that the Hamiltonian function has a unique minimum for real $x,y$, and one concludes that it is highly likely  the perturbative series $\mathcal E^{(\nu)}(\hbar)$ is Borel summable.
	
	The positions of the poles are related to the asymptotic behavior of the coefficients $E_n^{(k)}$. The large order factorial growth of the coefficients $\mathcal E^{(\nu)}_n$ is expected to be dictated by the saddles of the phase-space functional associated with the partition function of the difference operator. Indeed preliminary studies of the model of local $\bF_0$ indicate that $E^{(0)}_n (-1)^n \sim n!/(2|S|)$, where $S$ is the action of a complex instanton tunneling from the minimum at $x=p=0$ to one of the closest complex minima (say $p=0,x=2\pi \ri$). On the other hand, generic cases are complicated by the fact that the instanton actions are complex (in the local $\bF_0$ model the leading instanton action is real and negative). We leave detailed studies of this kind for the future.

	\begin{table}
		\centering
		\resizebox{\linewidth}{!}{\begin{tabular}{>{$}c<{$} *{3}{>{$}l<{$}}}\toprule
			\text{order} & \hbar = \pi & \hbar = 2\pi & \hbar = 11 \pi/7\\\midrule
			40  & \underline{2.15491639958596}48455449184602 & \underline{2.881815429}880211319432 & \underline{2.57475086894}731333042995\\
			70  & \underline{2.15491639958596599731}28390136 & \underline{2.88181542992629}4396204 & \underline{2.5747508689489039}3702545\\
			100 & \underline{2.1549163995859659973135074}608 & \underline{2.881815429926296782}625 & \underline{2.57475086894890395737}344\\\midrule
			\text{num.} & 2.1549163995859659973135074591 & 2.881815429926296782477 & 2.57475086894890395737295 \\\bottomrule
		\end{tabular}}
		\caption{The Borel-Pad\'{e} sums of the perturbative ground state energy $\mathcal E^{(0)}$ of the local $\bF_0$ with $m_1 = 1$ with various orders of truncation, compared with the stable numerical results. Underlined are the digits of the Borel-Pad\'{e} sums which are identical with the numerical results.}\label{tb:F2-detail}
	\end{table}

\begin{figure}
		\centering
		\subfloat[$F_1$]{\includegraphics[width=\figsize]{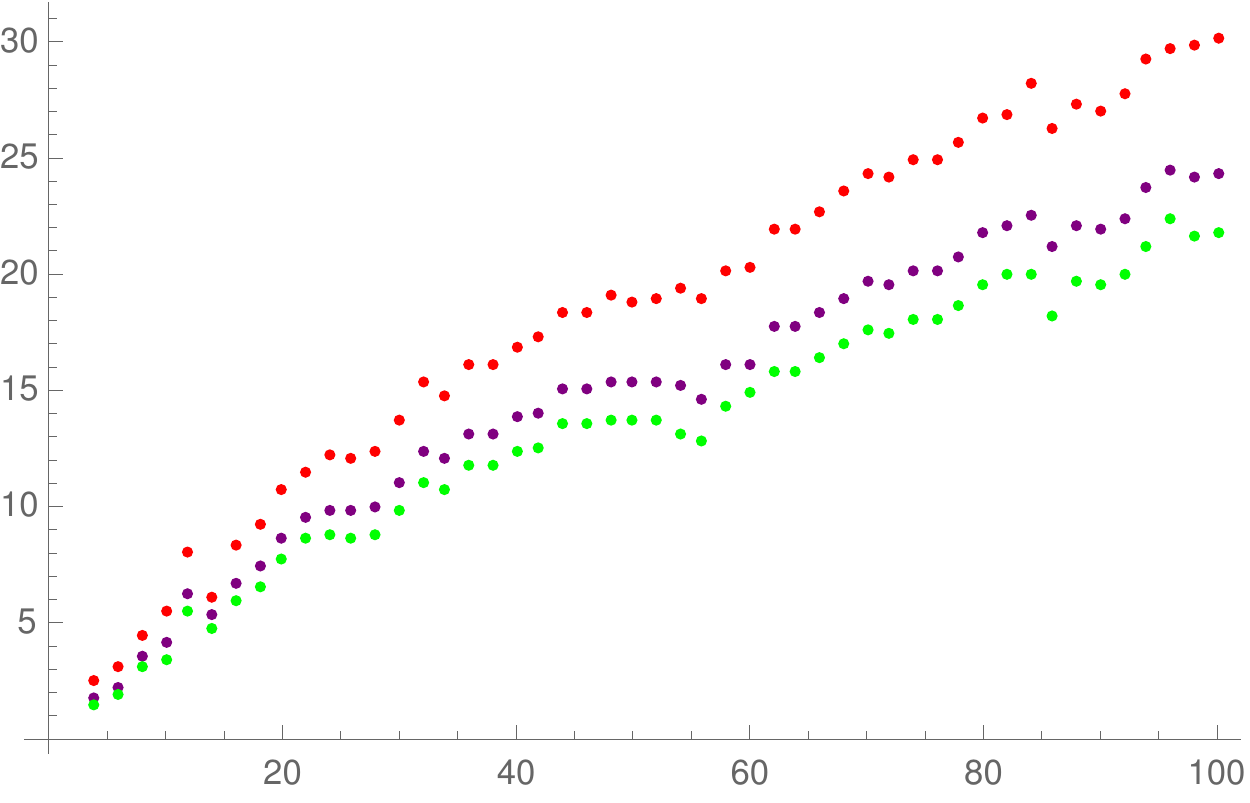}}\hspace{6ex}
		\subfloat[$F_2$\label{fg:md-F2}]{\includegraphics[width=\figsize]{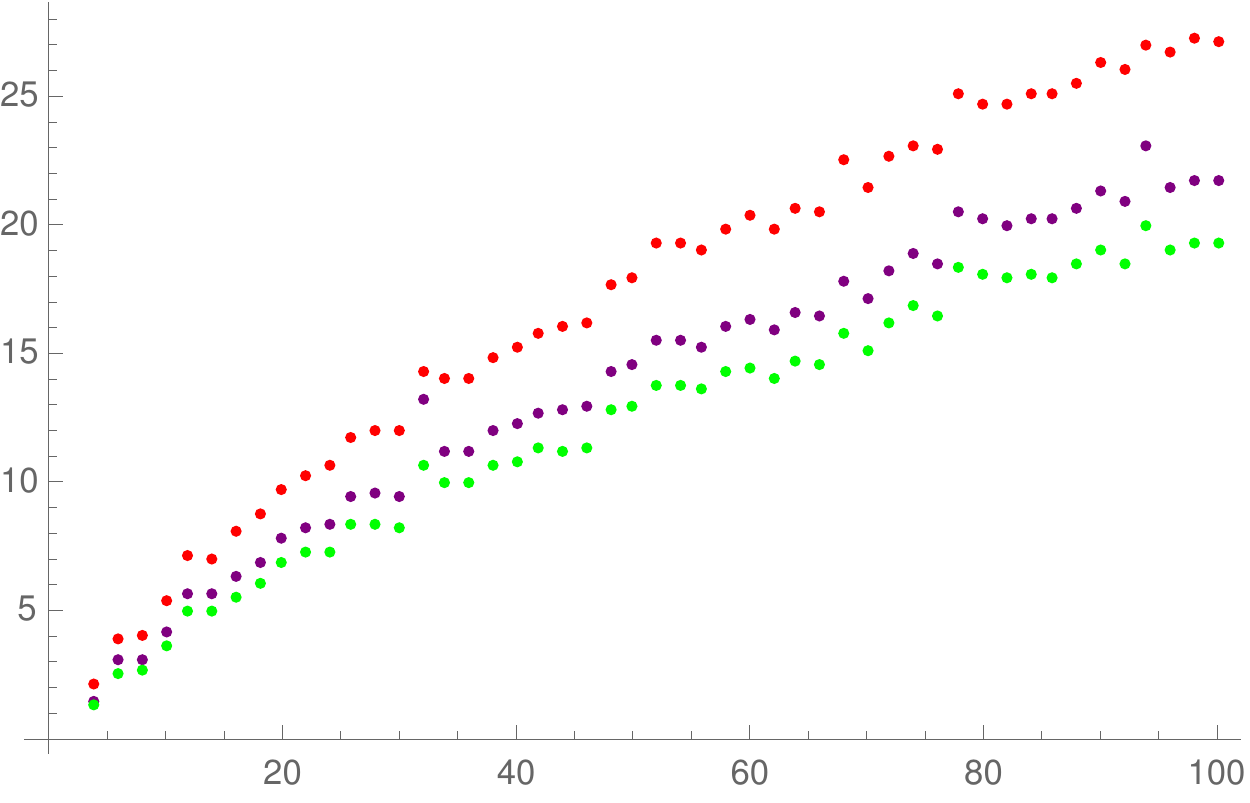}}\\
		\subfloat[$F_3$]{\includegraphics[width=\figsize]{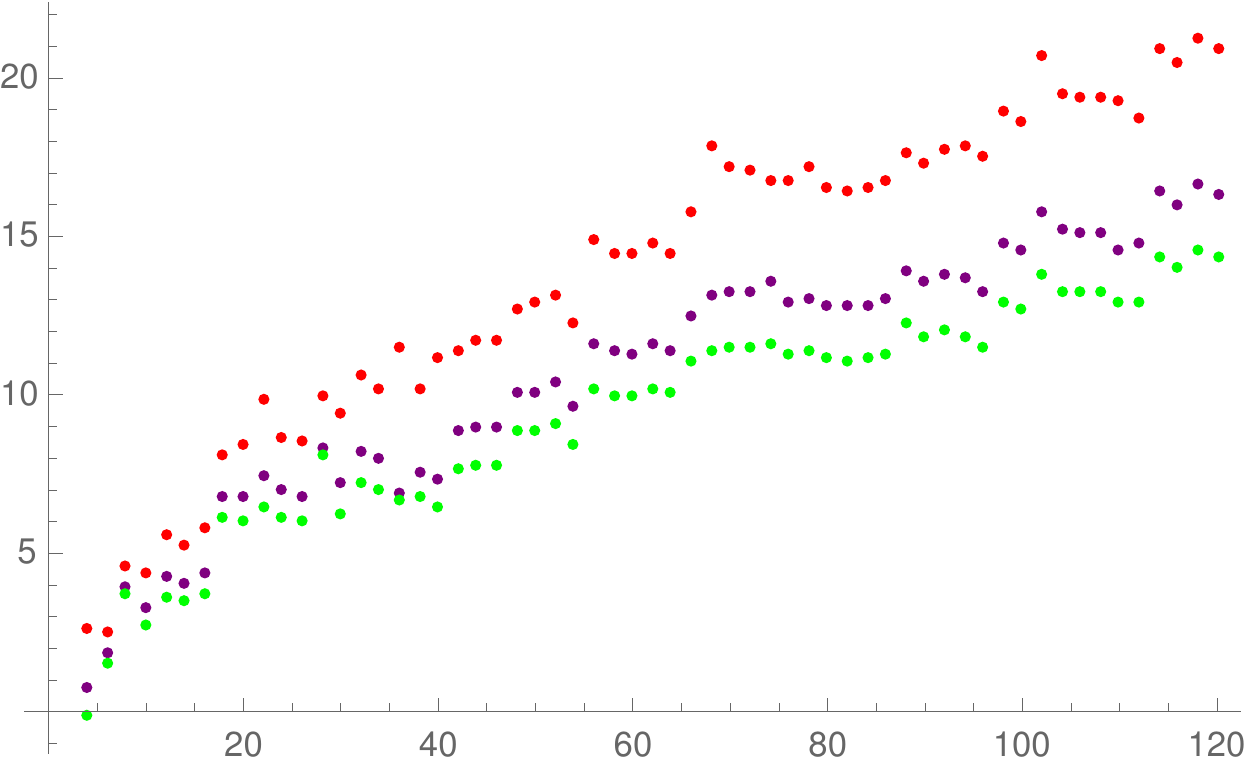}}\hspace{6ex}
		\subfloat[$F_4$]{\includegraphics[width=\figsize]{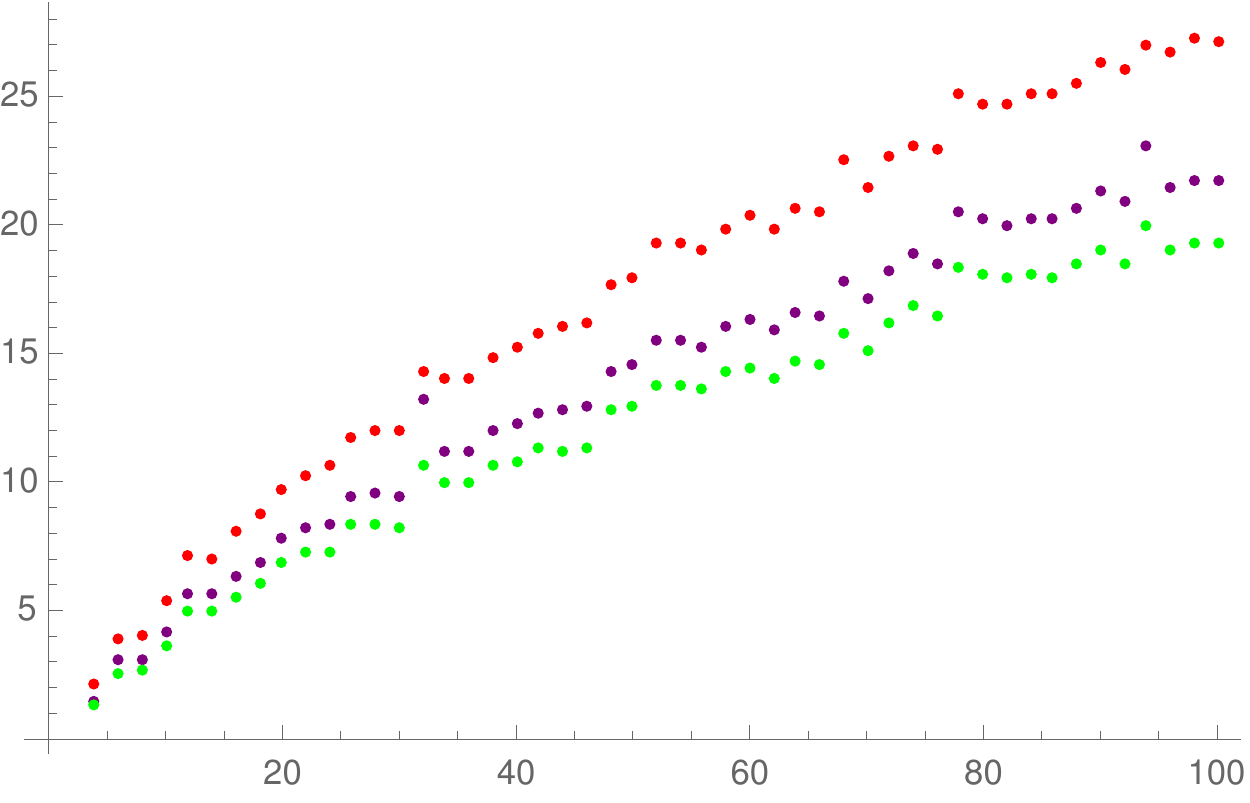}}\\
		\subfloat[$F_5$]{\includegraphics[width=\figsize]{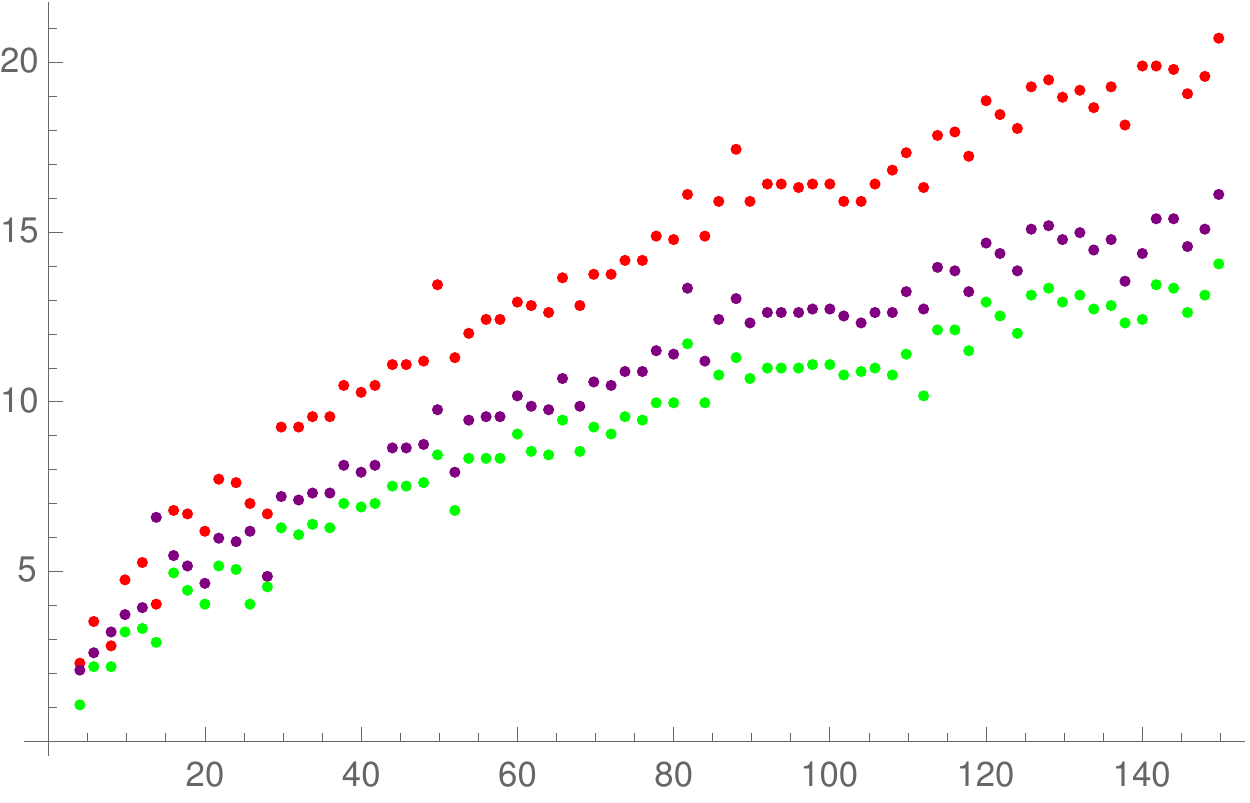}}\hspace{6ex}
		\subfloat[$F_6$]{\includegraphics[width=\figsize]{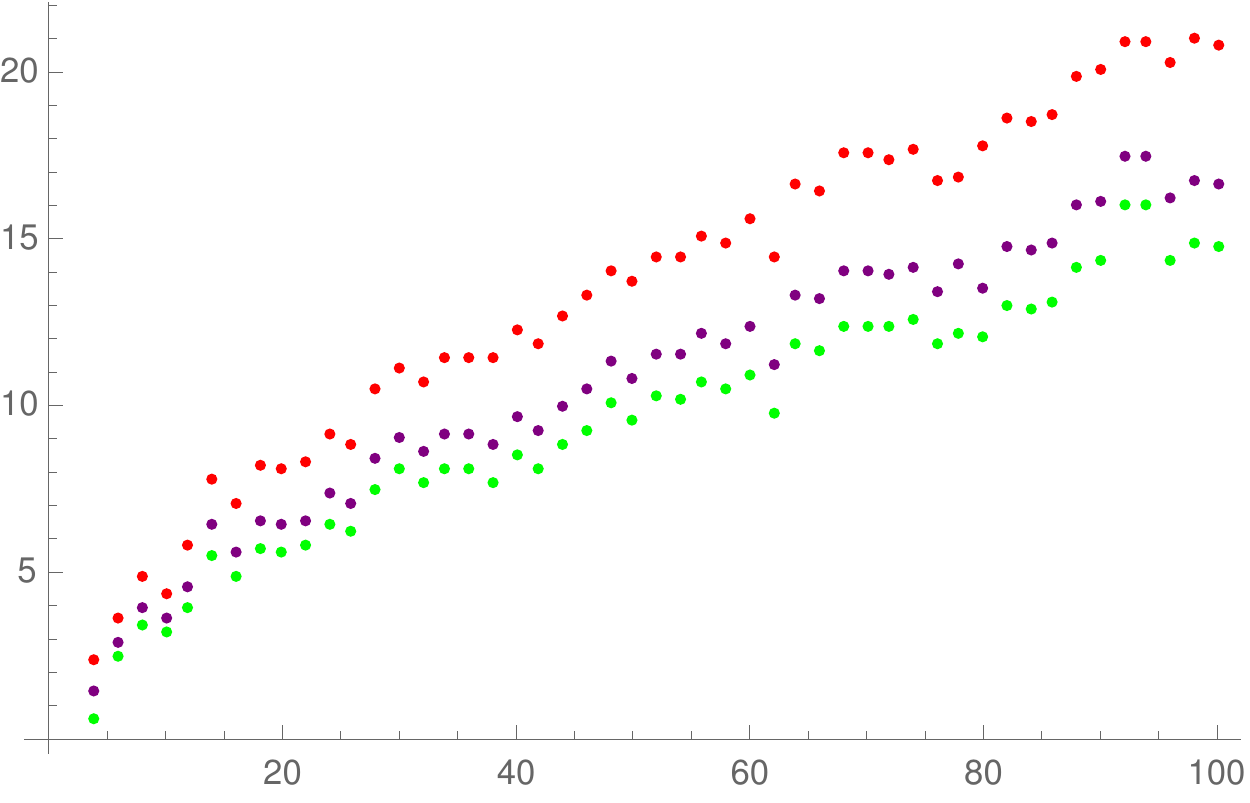}}\\
		\subfloat[$F_7$]{\includegraphics[width=\figsize]{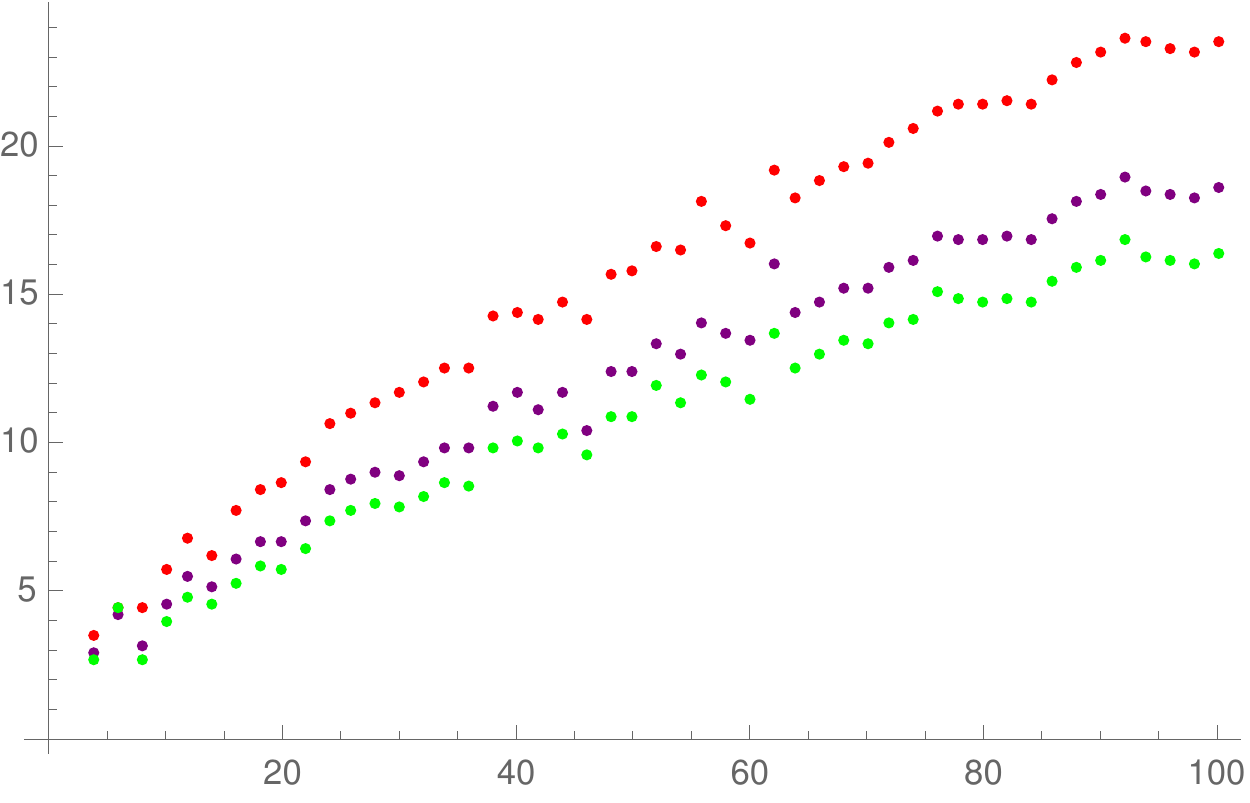}}\hspace{6ex}
		\subfloat[$F_8$]{\includegraphics[width=\figsize]{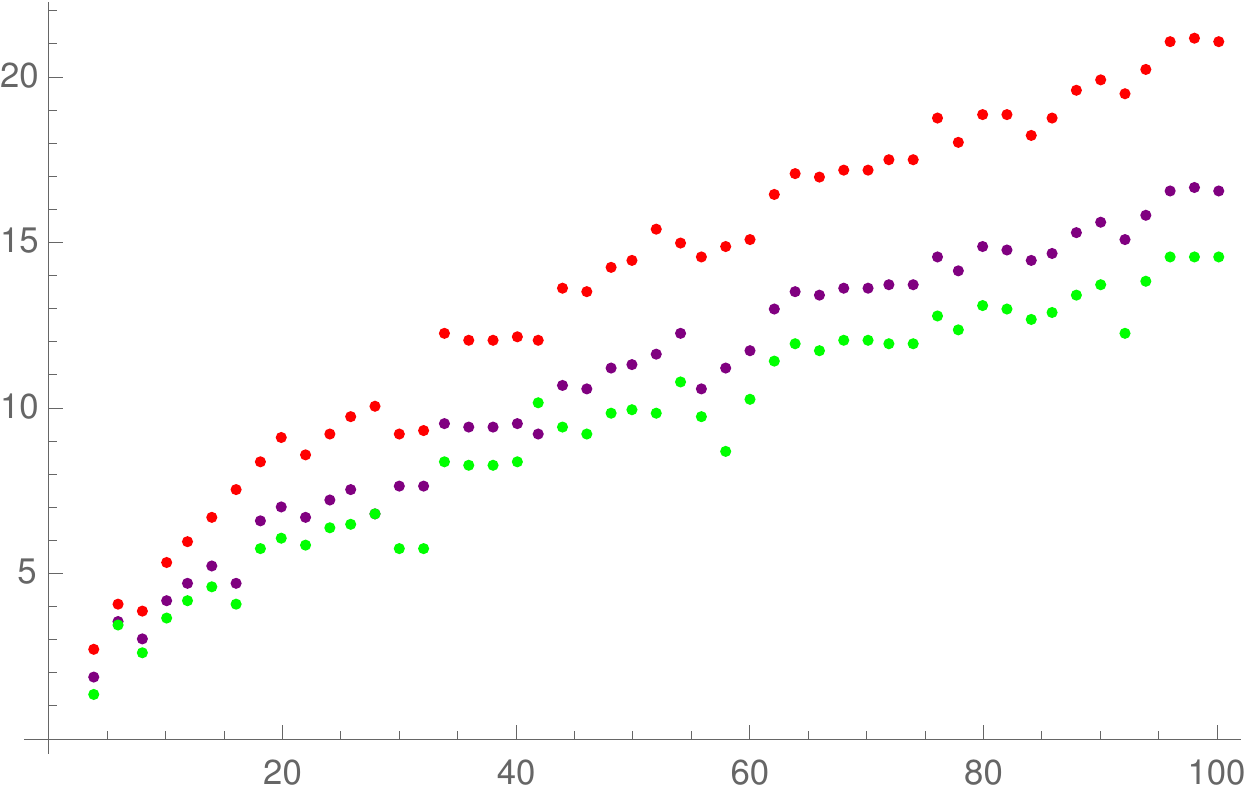}}		
		\caption{Plots of numbers of identical digits (measured by matching degree defined in \eqref{eq:md}) between Borel-Pad\'{e} sums and numerical results against orders of perturbative series for $F_1, \ldots, F_6$ with $\hbar = \pi$ (red), $\hbar = 11\pi/7$ (purple), and $\hbar = 2\pi$ (green). The increasing trend in these plots is a strong indication of the Borel summability of the spectrum.}\label{fg:md-1}
	\end{figure}

	We proceed to compute the Borel-Pad\'{e} sums of the perturbative ground state energy, evaluate them at $\hbar = \pi, 2\pi$, and $11\pi/7$, and compare with the numerical results. As seen in Tab.~\ref{tb:F2-detail}, both sides agree extremely well: the column of $\hbar = \pi$ agrees to 26 identical digits when 100 orders of $\hbar$ are taken. To better illustrate the success of the Borel-Pad\'{e} resummation, we define the matching degree between two numbers $x_1, x_2$
	\begin{equation}\label{eq:md}
		d(x_1, x_2) = -\log_{10}\left| \frac{x_1}{x_2} - 1 \right| \ ,
	\end{equation}
	which roughly speaking gives the number of identical digits between the two. We plot in Fig.~\ref{fg:md-F2} the matching degree between the Borel-Pad\'{e} sum and the numerical result against the truncation order of the perturbative series. It is very satisfactory to see that the matching degree grows up consistently with the perturbation order up to a very high value.

	\begin{table}
		\centering
		\resizebox{\linewidth}{!}{\begin{tabular}{*{1}{>{$}c<{$}} *{3}{>{$}l<{$}} c}\toprule
			\text{models} & \hbar = \pi & \hbar = 2\pi & \hbar = 11\pi/7 & order\\\midrule
			F_1     & 1.88885312929110349934403550512  & 2.56264206862381937081 & 2.28228027647413480906975 & 100\\
			F_2     & 2.1549163995859659973135074 & 2.881815429926296782 & 2.57475086894890395737 & 100\\
			F_3     & 2.74101669717594243806 & 3.3927922195048 & 3.112100386082561 & 120\\
			F_4     & 2.1549163995859659973135074 & 2.881815429926296782 & 2.57475086894890395737 & 100 \\
			F_5     & 2.850113139905259687 & 3.634196540335 & 3.30016753794720 & 150 \\
			F_6     & 2.4073757636270371349 & 3.24006352538625 & 2.888601794430404 & 100 \\
			F_7     & 2.6978665638653729660730 & 3.597651612809098 & 3.21315711223810717 & 100 \\
			F_8     & 2.50138703088653563645 & 3.39255629294437 & 3.0112549349998660 & 100 \\
			F_9     & 2.6978665638653729660730 & 3.597651612809098 & 3.21315711223810717 & 100 \\
			F_{10} & 2.5058190837155466420 & 3.4580019916041 & 3.054589829399109 & 100 \\
			F_{11} & 2.6164661244154612 & 3.65403010865 & 3.210251444588 & 150 \\
			F_{12} & 3.2257191850930277499 & 4.2098442497572 & 3.785161831282169 & 140 \\
			F_{13} & 3.1191905717052696024792 & 4.41710867528169 & 3.86437075506602184 & 140 \\
			F_{14} & 3.45068437001909478426792 & 4.654856221339859 & 4.13566326073216628 & 120 \\
			F_{15} & 3.2995079539638215478335633 & 4.5447897991133861 & 4.010681919079852304 & 140 \\
			F_{16} & 3.6584971507031114577 & 4.99018215393 & 4.4078476317280 & 170 \\
			\bottomrule
		\end{tabular}}
		\caption{Borel-Pad\'{e} sums of the perturbative ground state energies for the models listed in Tab.~\ref{tb:models}. The presented digits are both stable and identical with numerical results. The last column gives the orders of perturbative series used in the Borel-Pad\'{e} sums. }\label{tb:numericals}
	\end{table}	
	
\begin{figure}
		\centering
		\subfloat[$F_9$]{\includegraphics[width=\figsize]{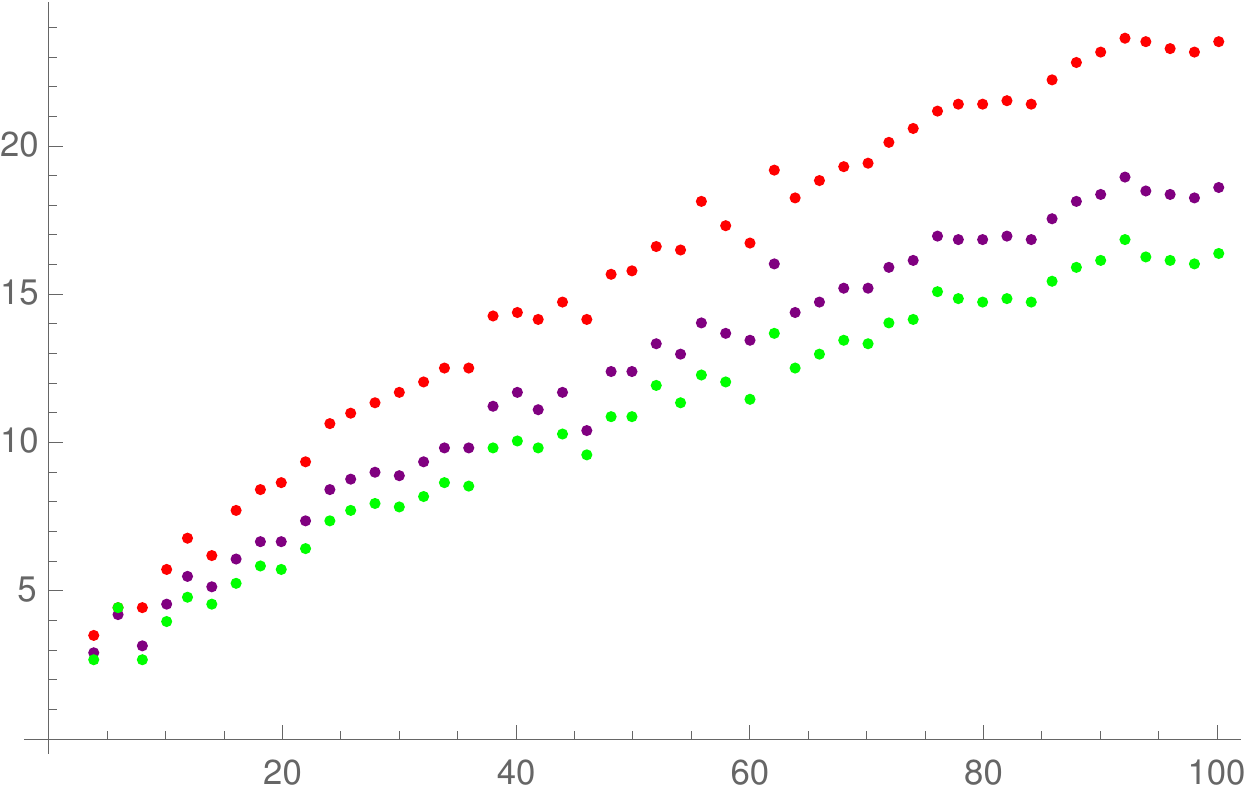}}\hspace{6ex}
		\subfloat[$F_{10}$]{\includegraphics[width=\figsize]{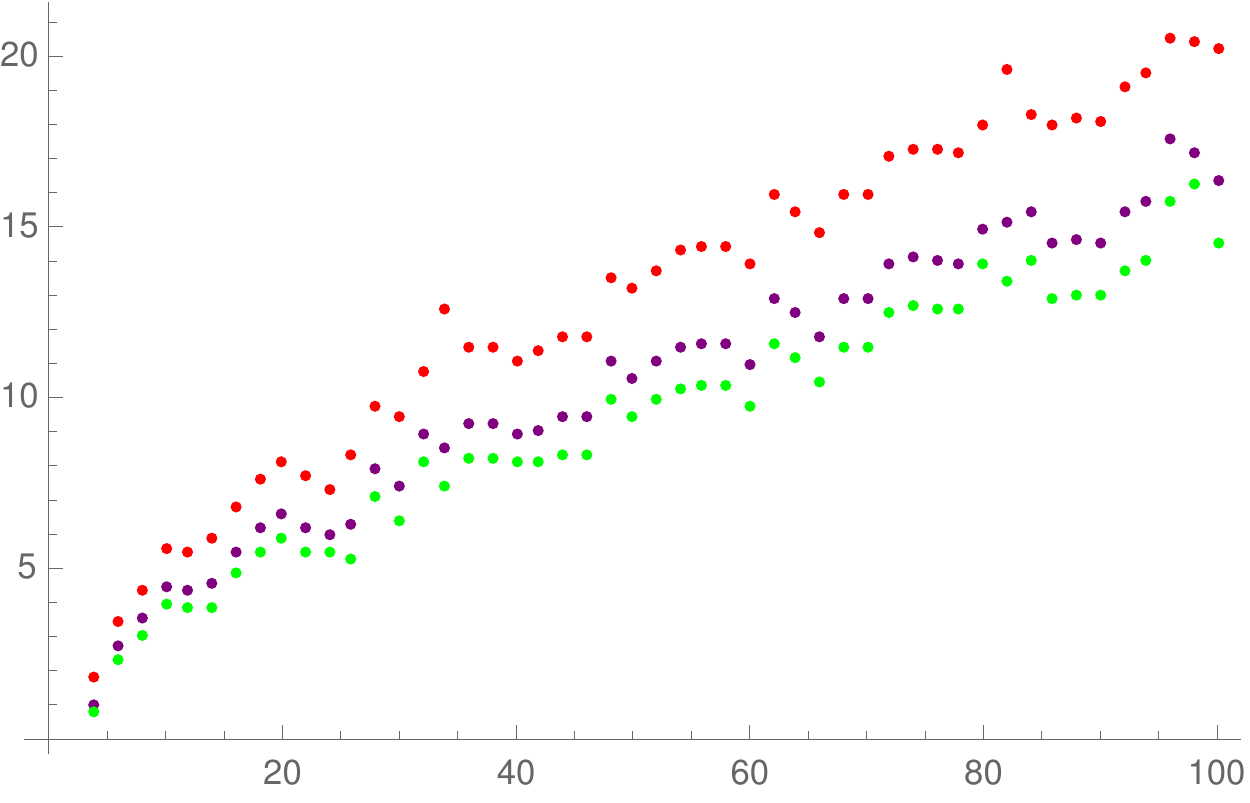}}\\
		\subfloat[$F_{11}$]{\includegraphics[width=\figsize]{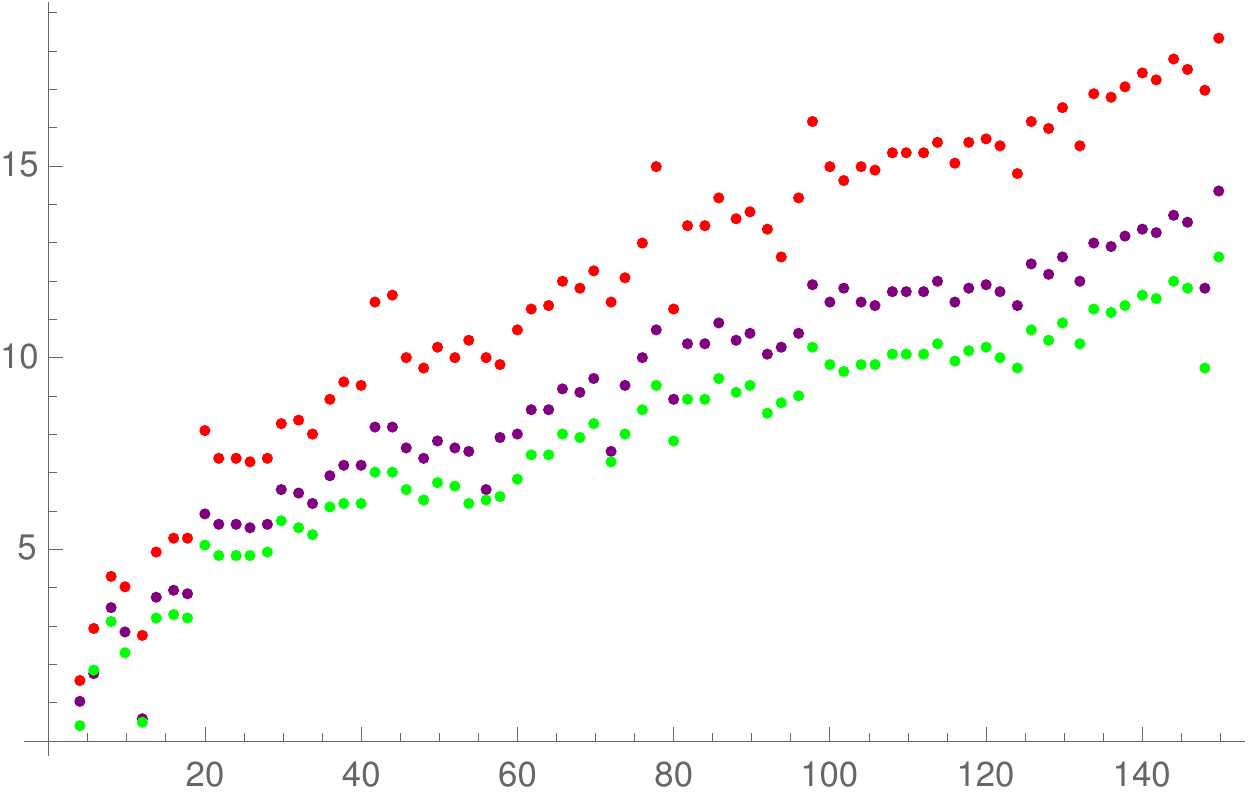}}\hspace{6ex}
		\subfloat[$F_{12}$]{\includegraphics[width=\figsize]{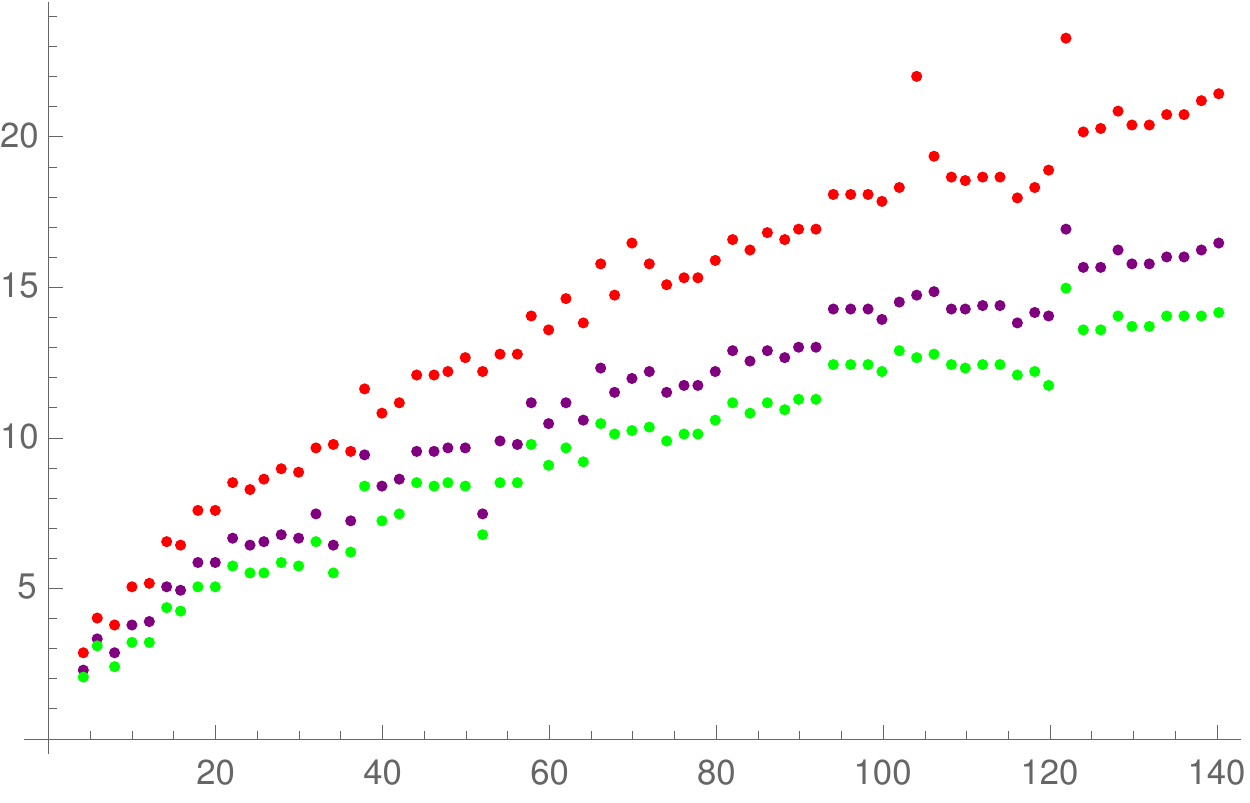}}\\
		\subfloat[$F_{13}$]{\includegraphics[width=\figsize]{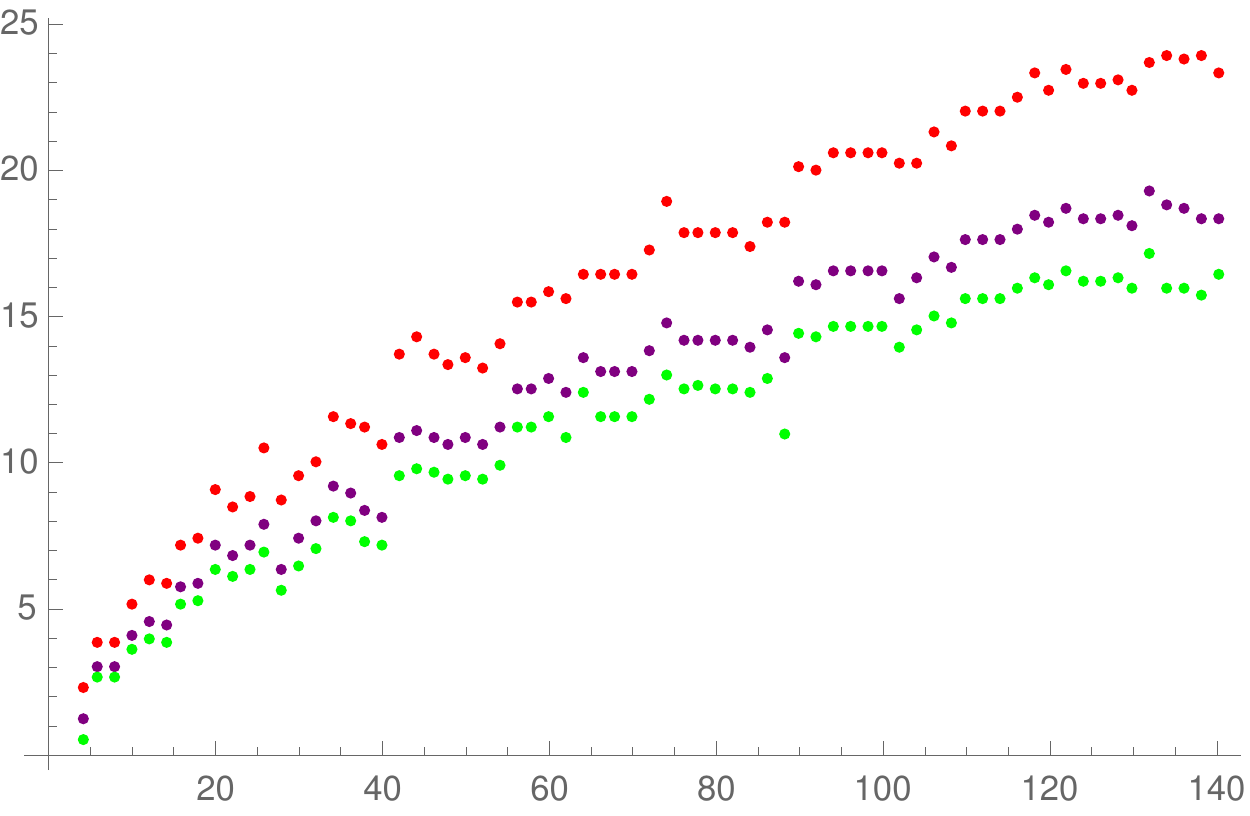}}\hspace{6ex}
		\subfloat[$F_{14}$]{\includegraphics[width=\figsize]{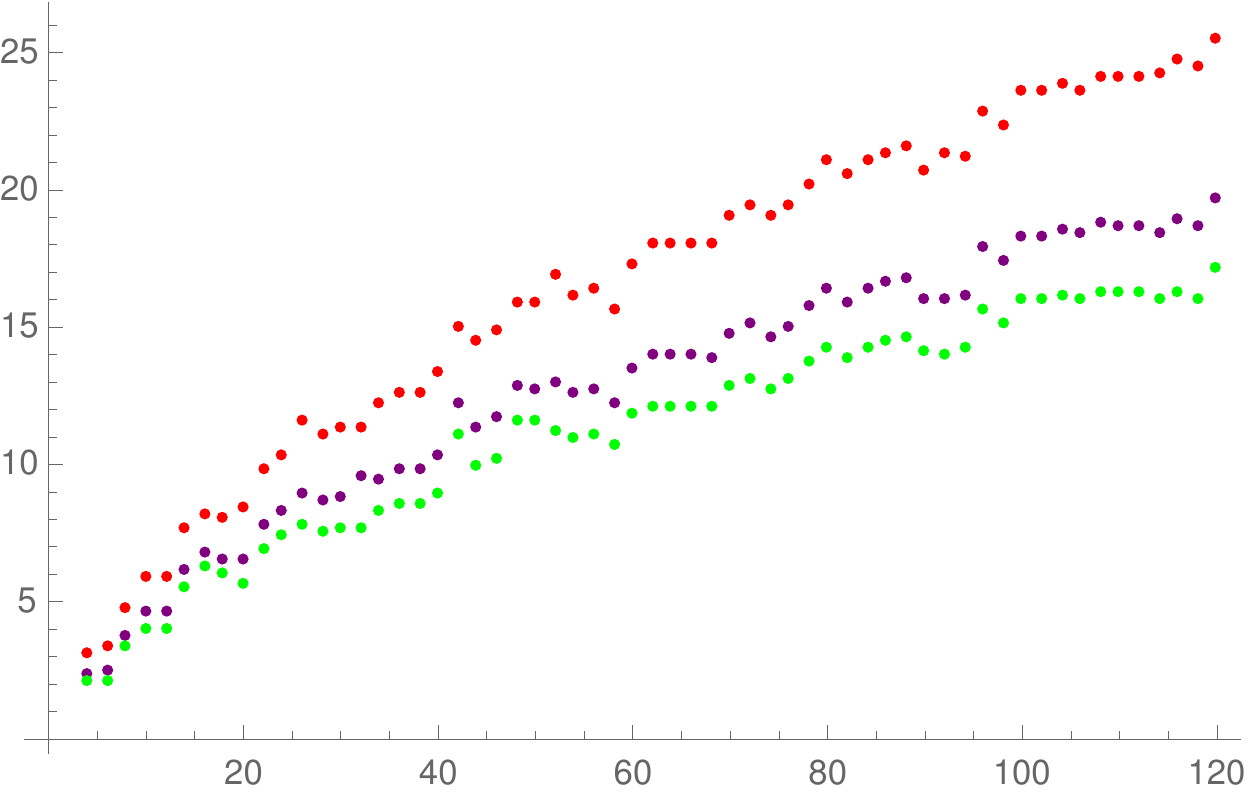}}\\
		\subfloat[$F_{15}$]{\includegraphics[width=\figsize]{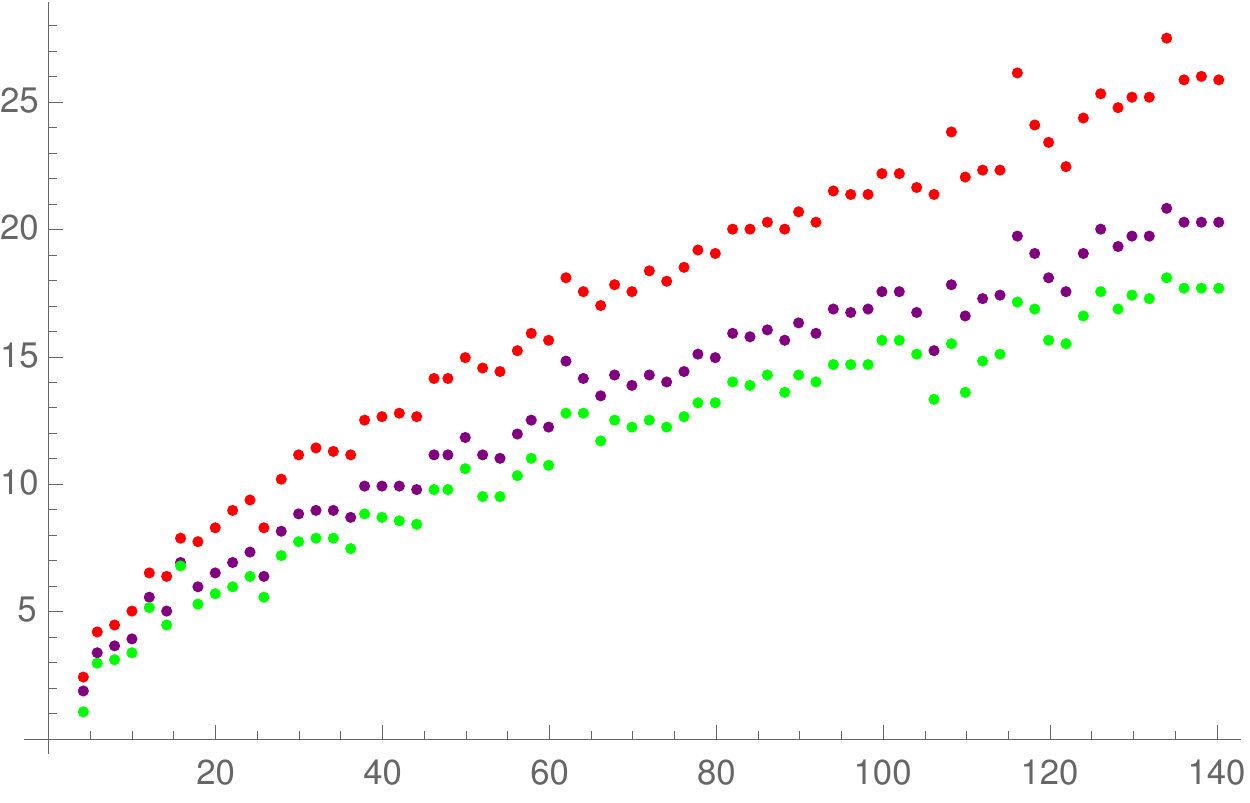}}\hspace{6ex}
		\subfloat[$F_{16}$]{\includegraphics[width=\figsize]{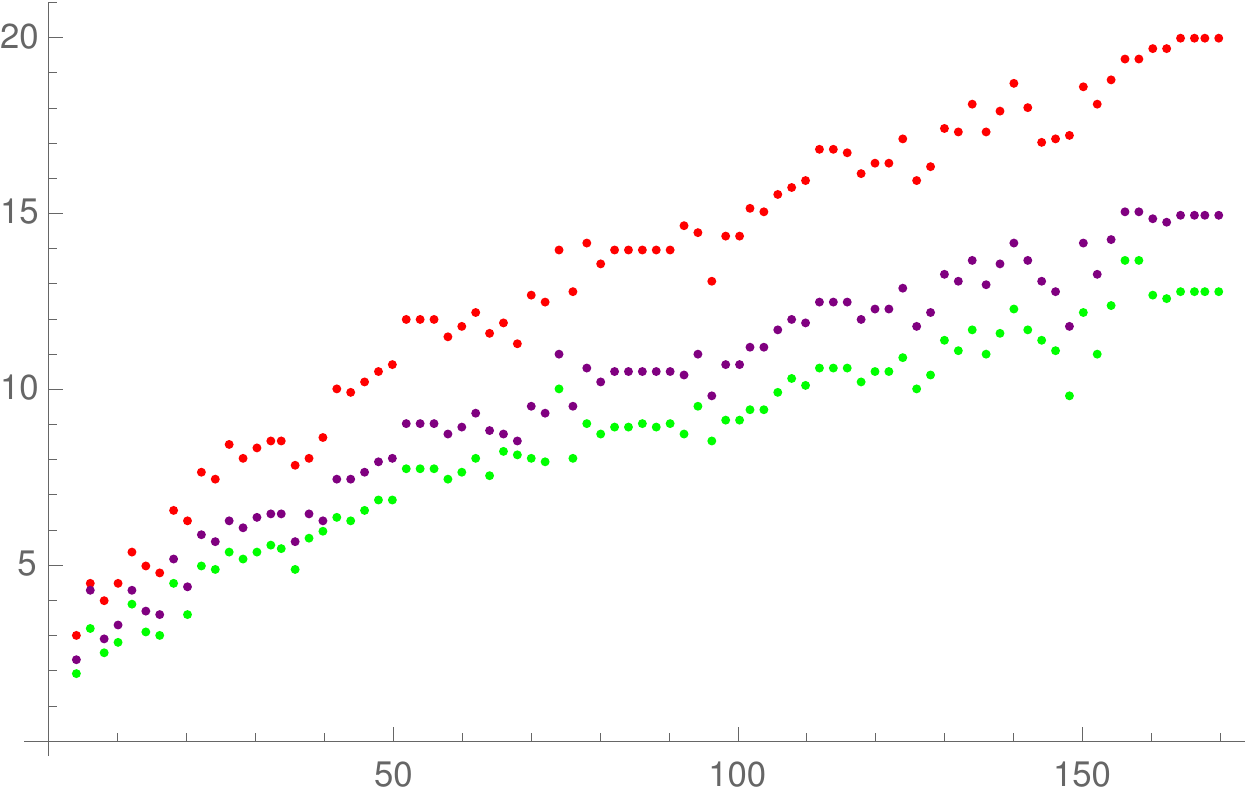}}		
		\caption{Same as in Fig.~\ref{fg:md-1} for
	 $F_9, \ldots, F_{16}$ with $\hbar = \pi$ (red), $\hbar = 11\pi/7$ (purple), and $\hbar = 2\pi$ (green).}\label{fg:md-2}
	\end{figure}

	We perform the same analysis for the other 15 models listed in Tab.~\ref{tb:models}. We find that in all 15 models, there are no stable poles along the positive real axis in the Borel plane, and we find agreement between the Borel-Pad\'{e} sums of the perturbative ground state energy and the numerical results, the degree of which improves consistently with increasing truncation order of the perturbative series. The plots of matching degrees for all 15 models are given in Figs.~\ref{fg:md-1}, \ref{fg:md-2}. Finally, we give in Tab.~\ref{tb:numericals} for all models the digits of the Borel-Pad\'{e} sums which are both stabilised and identical with the numerical results.
	
	We mention in passing that the underlying reason for the Borel summability of the spectrum is likely a consequence of the fact that no real-positive action instanton solutions exist in the limit of $\hbar\rightarrow 0$\footnote{The reason for this is that the Hamiltonian operators have a unique minimum as a function of $\sx$ and $\sp$ (see Appendix \ref{sc:uniqueness}).}. To show this one would need to carefully study the stokes phenomena as the phase of $\hbar$ is varied. We leave it as an open problem for the future.	
	
	
	\section{Conclusions and future prospects}
	\label{sc:cons}

	In this paper, we are interested in solutions to the eigenvalue problem of Hamiltonian operators which are difference operators, and in particular polynomials of $\re^\sx, \re^\sp$ with the commutation relation $[\sx, \sp] = \ri \hbar$. We developed Bender-Wu like recursion relations that solve eigenenergies and wave-functions perturbatively in small $\hbar$, and implemented the algorithm for \Mathematica{} in a function called \BWDifference{} in the updated \BenderWu{} package, originally developed in \cite{Sulejmanpasic:2016fwr}. Our algorithm is very efficient, capable of computing more than one hundred orders of perturbative solutions for a typical Hamiltonian difference operator in a reasonable amount of time.
	
	Typical Hamiltonian difference operators appear in the quantisation of mirror curves in topological string theory on toric Calabi-Yau threefolds. We studied all sixteen toric fano Calabi-Yau threefolds, whose associated Hamiltonian operators are unique, and computed the perturbative ground state energies for some choice of mass parameters. We find strong evidence that the perturbative eigenenergies are Borel summable, and the Borel sums are exact. Although we only studied and presented explicitly ground states, one can easily check the situation is the same in excited states. A possible reason for Borel summability is that the Hamiltonian difference operators arising in mirror curve quantization all have a unique real minimum as a function of $x$ and $p$, so that that classical equations of motion do not allow for real-positive-action instantons. However Borel summability (or even convergence) does not always mean that the re-summation gives the correct result, and non-perturbative corrections may still arise  (see e.g.~\cite{Grassi:2014cla,Dunne:2016jsr, Kozcaz:2016wvy}). Nevertheless the perturbation theory is factorially growing, and is likely dictated by the complex instanton (or ghost instanton) solutions which are generically present in such systems. Our \BWDifference{} function can be used to address such features in detail. 
	
	In addition, the mirror curves associated to toric fano Calabi-Yau threefolds are all genus-1 curves. In quantum mechanical systems with genus-1 curves obey a remarkable relation -- the {\'A}lvarez-Casares relation (\cite{alvarez2000-cubic,alvarez2000-generic,alvarez2002-quartic,alvarez2004-doublewell}, more examples are later found in \cite{Dunne:2013ada,Dunne:2014bca,Dunne:2016qix,Misumi:2015dua,Dunne:2016jsr,Kozcaz:2016wvy,Codesido:2016dld}) -- between the trivial perturbation theory and perturbation theory around instantons. Similar relation should exist for the ghost instantons of the quantum mirror curves. However in this case, as we argued, such nonperturbative objects do not contribute in the trans-series expansion, but rather dictate the asymptotic growth. Because of this, we expect a form of self-resurgence \cite{Kozcaz:2016wvy} to hold, if the analogous {\'A}lvarez-Casares relation holds in quantum mirror curves.
	
	In this paper, we restrict ourselves to quantum mirror curves of fano Calabi-Yau threefolds. It would also be interesting to look at more generic toric Calabi-Yau threefolds, whose mirror curves have genera $g$ great than one. There would be two different but related quantum mechanical problems. The first is again the quantisation of mirror curves. The associated Hamiltonian operators are no longer unique \cite{Codesido:2015dia}, but each of them still gives rise to a one-dimensional quantum mechanical system. Alternatively, one could also look at the cluster integral systems \cite{Goncharov2011} associated to the toric Calabi-Yau threefolds. When the latter are not fano, the cluster integral systems are higher dimensional, involving $g$ mutually commutative Hamiltonian operators, all of which are difference operators of the exponential-polynomial type. In order to study perturbatively the cluster integral systems, we would need to generalise our algorithm to treat multivariable systems. In addition, it would be desirable to further generalise the Bender-Wu algorithm for more generic difference operators, not necessarily of the exponential-difference type. Another interesting question is whether our Bender-Wu solutions to the Hamiltonian difference operators, assuming wave-functions can be expanded in terms of the wave-functions of harmonic oscillators, exhaust all possible wave-functions in the domain $\cD$ of the Hamiltonian difference operators.

%
	
	\section*{Acknowledgement}
	
	We would like to thank Gerald Dunne, Alba Grassi, Yasuyuki Hatsuda, Amir-Kian Kashani-Poor, Albrecht Klemm, Marcos Mari\~{n}o, and Mithat \"Unsal for useful discussions. We appreciate the careful reading of the manuscript by Alba Grassi and Yasuyuki Hatsuda, and thank Gerald Dunne and Marcos Mari\~{n}o for pointing out some important references. J.G. is supported by the grant  ANR-13-BS05-0001.

	\clearpage
	\begin{appendix}

	\section{The Bender-Wu recursion relations}
	\label{sc:BW-rec}
	
	We derive in detail here the solution to the following eigenvalue problem with Bender-Wu type recursion relations,
	\begin{equation}
		\cH(\sqrt{\hbar}\,\hx, \sqrt{\hbar}\,\hp) \Psi(x) = E \Psi(x) \ ,
	\end{equation}
	where
	\begin{equation}
		\cH(\sqrt{\hbar}\,\hx, \sqrt{\hbar}\,\hp) = \sum_{r,s} a_{r,s} \re^{\sqrt{\hbar}(r\hx + s\hp)} \ ,
	\end{equation}
	and $\hx, \hp$ satisfy the commutation relation
	\begin{equation}
		[\hx, \hp] = \ri \ .
	\end{equation}
	
	The Hamiltonian $\cH(\sqrt{\hbar}\,\hx, \sqrt{\hbar}\,\hp)$ is assumed to have no linear term in $\hx,\hp$ in small $\hbar$ expansion. Therefore
%
%
\be
\cH(x\sqrt\hbar,p\sqrt{\hbar})=\sum_{r,s} a_{r,s} + \frac{\hbar}{2}(A \hx^2+B\hp^2+C( \hx \hp + \hp \hx))+\cO(\hbar^{3/2})\;,
\ee
where
\be\label{eq:ABC}
A=\sum_{r,s}r^2 a_{r,s}\;,\qquad B=\sum_{r,s} s^2 a_{r,s}\;,\qquad C=\sum_{r,s} r s\:a_{r,s}\;.
\ee
We wish that in the limit $\hbar\rightarrow 0$ the Hamiltonian reduces to a harmonic oscillator. Therefore we perform the canonical transformation
\begin{equation}
	(\hx, \hp) \mapsto (\xi\hx, \xi^{-1}\hp + \alpha \xi \hx) \ ,
\end{equation}
with
\be\label{eq:alpha-xi}
\alpha=-C/B \ ,\quad \xi=\left(\frac{B^2}{AB-C^2}\right)^{1/4}\ .
\ee
so that to lowest orders the Hamiltonian operator becomes a simple harmonic oscillator with unit mass and frequency
\be
\text{const.} + \frac{\hbar}{2}\sqrt{AB-C^2}\left(\hx^2+\hp^2\right)\;.
\ee
Hence we can define a reduced Hamiltonian 
\begin{equation}
h\(\sqrt\hbar\,\hx, \sqrt\hbar\,\hp\)=\frac{1}{(1/2)\sqrt{AB-C^2}}\cH\(\sqrt\hbar\,\xi \hx , \sqrt{\hbar}\,(\xi^{-1} \hp+\alpha\xi \hx)\) \ .
\end{equation}
Now we wish to solve the eigenvalue equation\footnote{Notice that if $h$ is invariant under $p\rightarrow -p$, then for every solution $\psi(x)$ we have that $\psi^*(x)$ is also a solution. This means that we can always choose a real solution. This in turn will guarantee that all the $A$ and $\tilde A$-coefficient appearing below can be made real with the apropreate choice of normalization.}
\begin{equation}\label{eq:h-eq}
	h\( \sqrt{\hbar}\,\hx, \sqrt{\hbar}\,\hp \) \psi(x) = \hbar \epsilon \psi(x) \ .
\end{equation}
The eigenvalues and wave-functions are related to those of $\cH\(\sqrt{\hbar}\,\hx, \sqrt{\hbar}\,\hp\)$ by
\begin{equation}
	E = \frac{\hbar}{2}\sqrt{AB-C^2} \epsilon \ ,\quad \Psi(x) = \re^{\ri\frac{\alpha}{2}x^2} \psi(x/\xi) \ .
\end{equation}
We also comment here that although $h(\sqrt{\hbar}\,\hx, \sqrt{\hbar}\,\hp)$ is a difference operator, at any finite order in $\sqrt\hbar$ expansion it is a polynomial in $\hx, \hp$ and thus a finite order differential operator.

%
%

In order to solve the eigenvalue problem \eqref{eq:h-eq}, it is convenient to write the coordinate $\hx$ and momentum $\hp$ operators in terms of the creation and the annihilation operators 
\be
\hx=\frac{1}{\sqrt{2}}(a^\dagger+a)\;,\qquad \hp=\frac{\ri}{\sqrt 2}(a^\dagger-a)\;.
\ee
The operator $h$ becomes
\be\label{eq:h_def}
\frac{h}{\hbar}=\frac{1}{g^2}\sum_{r,s}\tilde a_{r,s}e^{g\beta(r,s) a^\dagger+g\bar\beta(r,s) a}\;,
\ee
where we labeled
\be
g=\sqrt{\hbar/2}\;,\qquad \beta(r,s)=(\alpha s+r)\xi+is/\xi\;,\qquad \tilde a_{r,s}=\frac{1}{\sqrt{AB-C^2}} a_{r,s} \ .
\ee
It can be checked by explicit calculations that
\begin{equation}\label{eq:beta-ids}
	\sum_{r,s} \tilde{a}_{r,s} \beta(r,s) = \sum_{r,s} \tilde{a}_{r,s} \beta(r,s)^2 = 0\ ,\quad \sum_{r,s} \tilde{a}_{r,s} |\beta(r,s)|^2 = 2 \ .
\end{equation}
It is beneficial to normal order the reduced Hamiltonian by writing all the annihilation operators to the right of the creation operators. Using the BCH identity
we can write $h$ as
\be
\frac{h}{\hbar}=\frac{1}{g^2}\sum_{r,s} \tilde a_{r,s}e^{g\beta a^\dagger}e^{g\bar\beta a}e^{|\beta|^2\frac{1}{2}g^2}
\ee
or, by expanding the exponents
\be
\frac{h}{\hbar}=\sum_{r,s}\sum_{n_1,n_2,n_3} \tilde{a}_{r,s} \frac{\beta^{n_1}\bar\beta^{n_2}|\beta|^{2n_3}}{n_1!n_2!n_3!}g^{n_1+n_2+2n_3-2}\frac{1}{2^{n_3}}(a^\dagger)^{n_1}a^{n_2}
\ee
where $n_1,n_2,n_3$ run from $0$ to infinity.

Now we make a formal expansion of the wave-function and eigenvalue
\be
\psi(x)=\sum_{l,k=0}^\infty A_l^kg^l \psi_k(x)\;,\qquad \epsilon=\sum_{l=0}^\infty\epsilon_{l-2}g^{l-2}\;,
\ee
where $\psi_k(x)$ are eigenfunctions of the simple harmonic oscillator with unit mass frequency (i.e. $\psi_\nu(x)$ are solutions of the leading order spectral problem). Using the fact that\footnote{Note that this expression vanishes if $n_2>k$, as it should, because the factorial function of negative numbers is infinite.}
\be
(a^{\dagger})^{n_1}a^{n_2}\psi_l=\frac{\sqrt{k!(k+n_1-n_2)!}}{(k-n_2)!}\psi_{k-n_2+n_1}\;,
\ee
we get
\be
\begin{aligned}
\sum_{n_1,n_2,n_3}\sum_{r,m}\tilde a_{r,s}\frac{\beta^{n_1}\bar\beta^{n_2}|\beta|^{2n_3}}{n_1!n_2!n_3!}\frac{1}{2^{n_3}}\frac{\sqrt{k!(k+n_1-n_2)!}}{(k-n_2)!}A_l^k&g^{l+n_1+n_2+2n_3-2}\psi_{k-n_2+n_1}\\
&=\sum_{l,n}\epsilon_{n-2}A_{l}^kg^{l+n-2}\psi_k\;.
\end{aligned}
\ee
By equationg powers of $g$ and coefficients of $\psi_k$ on both sides, we have
\be
\sum_{n_1,n_2,n_3}\sum_{r,m}\tilde a_{r,s}\frac{\beta^{n_1}\bar\beta^{n_2}|\beta|^{2n_3}}{n_1!n_2!n_3! 2^{n_3}}\frac{\sqrt{(k+n_2-n_1)!k!}}{(k-n_1)!}A_{l-n_1-n_2-2n_3}^{k+n_2-n_1}=\sum_n \epsilon_{n-2}A_{l-n}^k\;.
\ee
Notice that we can formally assume that $n_{1,2,3}$ run from $-\infty$ to $+\infty$, noting that the factorials have poles at negative integer values, and that $A_{l}^k$ vanishes for negative $k$ or $l$. Then we can freely shift $n_1\rightarrow n_1+n_2$ without worrying about the limits of the sum, and get
\be
\sum_{n_1,n_2,n_3}\sum_{r,m}\tilde a_{r,s}\frac{\beta^{n_1}|\beta|^{2(n_2+n_3)}}{(n_1+n_2)!n_2!n_3!}\frac{1}{2^{n_3}}\frac{\sqrt{(k-n_1)!k!}}{(k-n_1-n_2)!}A_{l-n_1-2(n_2+n_3)}^{k-n_1}=\sum_n \epsilon_{n-2}A_{l-n}^k\;.
\ee
Now we shift $n_3\rightarrow n_3-n_2$ to get
\be
\sum_{n_1,n_2,n_3}\sum_{r,s}\tilde a_{r,s}\frac{\beta^{n_1}|\beta|^{2n_3}}{(n_1+n_2)!n_2!(n_3-n_2)!}\frac{1}{2^{n_3-n_2}}\frac{\sqrt{(k-n_1)!k!}}{(k-n_1-n_2)!}A_{l-n_1-2n_3}^{k-n_1}=\sum_n \epsilon_{n-2}A_{l-n}^k\;.
\ee
Notice that the sum over $n_2$ can now be performed\footnote{This simply follows from the definition of the hypergeometric function
\[
F(a,b;c;z)=\sum_{s=0}^\infty \frac{1}{s!}\frac{(a)_s(b)_s}{(c)_s}z^s\;,
\]
where $(a)_s=(a)(a+1)\dots(a+s-1)=\frac{\Gamma(a+s)}{\Gamma(a)}$. If $a,b$ are negative integers $-n,-m$ then we can use the Gamma-function reflection formula to get that $(-n)_s=(-1)^s \frac{n!}{(n-s)!}$. Further if we take that $c=q+1$ with $q\in \mathbb N^0$, we have that $(q+1)_s=\frac{(q+s)!}{q!}$ so that
\[
F(-n,-m;q+1;z)=\sum_{s=0}^\infty \frac{q!n!m!}{s!(n-s)!(m-s)!(q+s)!}z^s\;,
\]
which gives
\[
\sum_{s=0}^\infty \frac{1}{s!(n-s)!(m-s)!(q+s)!}z^s=\frac{F(-n,-m;1+q;z)}{q!n!m!} \ .
\]}
\begin{equation}
\sum_{n_2=0}^\infty \frac{2^{n_2}}{(n_1+n_2)!n_2!(n_3-n_2)!(k-n_1-n_2)!}=\begin{cases}\frac{F(-k+n_1,-n_3;1+n_1;2)}{(k-n_1)!n_1!n_3!}&n_1\ge 0\\
\frac{F(-k,-n_1-n_3;1-n_1;2)}{2^{n_1}k!(-n_1)!(n_1+n_3)!} & n_1<0\;.
\end{cases}
\end{equation}
where $F(a,b;c;z)=_2F_{1}(a,b;c;z)$ is the hypergeometric function. 
Relabeling $n_3$ by $q$, $n_1$ by $n$ or $-n$ if $n_1$ is negative, we have that
\begin{multline}
\sum_{n\ge 0,q}\sum_{r,s}\tilde a_{r,s} \frac{\beta^{n}|\beta|^{2q}}{n! q!}\frac{1}{2^{q}}\sqrt{\frac{k!}{(k-n)!}}F(-k+n,-q;1+n;2)A_{l-n-2q}^{k-n}
\\+\sum_{n< 0,q}\sum_{r,s}\tilde a_{r,s} \frac{\beta^{-n}|\beta|^{2q}}{n! (q-n)!}\frac{1}{2^{q-n}}\sqrt{\frac{(k+n)!}{k!}}F(-k,n-q;1+n;2)A_{l+n-2q}^{k+n}=\sum_{n}\epsilon_{n-2}A_{l-n}^k\;.
\end{multline}
Finally we make the shift $q\rightarrow q+n$ in the second summation to make the expression above in a nicer form.
\begin{multline}
\sum_{q=0}^\infty\sum_{r,s}\tilde a_{r,s} \frac{|\beta|^{2q}}{q!}\frac{1}{2^{q}}F(-k,-q;1;2)A_{l-2q}^{k}
\\+\sum_{n=1}^\infty\sum_{q=0}^\infty\sum_{r,s}\tilde a_{r,s} \frac{|\beta|^{2q}}{n! q!}\frac{1}{2^{q}}\Bigg(\beta^n\sqrt{\frac{k!}{(k-n)!}}F(-k+n,-q;1+n;2)A_{l-n-2q}^{k-n}\\+\bar\beta^n\sqrt{\frac{(k+n)!}{k!}}F(-k,-q;1+n;2)A_{l-n-2q}^{k+n}\Bigg)=\sum_{n=0}^\infty\epsilon_{n-2}A_{l-n}^k\;,\label{eq:id-recursion}
\end{multline}
where on the l.h.s. the $n=0$ term has been singled out. This identity is valid for any $k,l\geq 0$.

Let us consider some examples for the identity \eqref{eq:id-recursion}. When $l=0$, only terms with $q=0,n=0$ contribute, and the identity reduces to
\be
\sum_{r,s}\tilde a_{r,s}A_0^k=\epsilon_{-2}A_0^k\;, \quad \forall k \in \bN_0 \ .
\ee
Given that not all $A_0^k$ can vanish, one finds the classical energy
\be
\epsilon_{-2}=\sum_{r,s}\tilde a_{r,s} \ .
\ee
When $l=1$, using the identity \eqref{eq:beta-ids} reduces to
%
\be
\epsilon_{-1}=0\;.
\ee

Next we consider \eqref{eq:id-recursion} when $l \geq 2$. Note that the the summand of the first summation when $q=0$ always cancels with the term proportional to $\epsilon_{-2}$ on the r.h.s., and that the summand of the second summation when $(n,q) = (1,0), (2,0)$ vanish due to the identities \eqref{eq:beta-ids}. Therefore \eqref{eq:id-recursion} becomes
\begin{multline}\label{eq:master}
(2k+1-\epsilon_0)A_{l}^{k}+\sum_{q=2}^{\lfloor\frac{l+2}{2}\rfloor}\sum_{r,s}\tilde a_{r,s} \frac{|\beta|^{2q}}{q!}\frac{1}{2^{q}}F(-k,-q;1;2)A_{l+2-2q}^{k}
\\+\sum_{q=0}^{\lfloor\frac{l+2}{2} \rfloor}\sum_{n=\max(1,3-2q)}^{l+2-2q}\sum_{r,s}\tilde a_{r,s} \frac{|\beta|^{2q}}{n! q!}\frac{1}{2^{q}}\Bigg(\beta^n\sqrt{\frac{k!}{(k-n)!}}F(-k+n,-q;1+n;2)A_{l+2-n-2q}^{k-n}\\+\bar\beta^n\sqrt{\frac{(k+n)!}{k!}}F(-k,-q;1+n;2)A_{l+2-n-2q}^{k+n}\Bigg)=\sum_{n=1}^l\epsilon_{n}A_{l-n}^k\;.
\end{multline}
Here we have shifted the index $l\rightarrow l+2$ on both sides, $n \rightarrow n+2$ on the r.h.s., and then singled out the terms proportional to $A_l^k$.
Now notice that the sums on both the left and right hand side contain only coefficients $A_{\tilde l}^k$ with $\tilde l<l$. So by inserting $l=0$ all that remains is
\be
(2k+1-\epsilon_0)A_0^k=0\;, \quad k\in \bN_0 \ .
\ee
Since not all $A_0^k$ vanish, this identity can only be true if for some nonnegative integer $\nu$
\begin{equation}
	\epsilon_0 = 2\nu+1, \quad A_0^\nu = \gamma \neq 0 \quad \text{and} \quad A_0^{k} = 0\ ,\; \forall k\neq \nu \ ,
\end{equation}
where $\gamma$ is an arbitrary nonvanishing constant. $\nu$ serves as the level of the eigenvalue/wave-function solution. Fixing the level $\nu$, we can normalise the wave-function so that
\begin{equation}
	A^\nu_l = 0\ ,\;\forall l > 0 \ .
\end{equation}

Following the normalisation of wave-function above, \eqref{eq:master} gives us two recursion relations that solve $A_l^k$ and $\epsilon_l$ respectively. Assuming that $A_{\tilde{l}}^k$ and $\epsilon_{\tilde{l}}$ are known for all $\tilde{l} < l$, the expansion coefficients $A_l^k$ and $\epsilon_l$ can be solved from
\begin{multline}
A_{l}^{k}=\frac{1}{2(k-\nu)}\Bigg(-\sum_{q=2}^{\lfloor\frac{l+2}{2}\rfloor}\sum_{r,s}\tilde a_{r,s} \frac{|\beta|^{2q}}{q!}\frac{1}{2^{q}}F(-k,-q;1;2)A_{l+2-2q}^{k}
\\-\sum_{q=0}^{\lfloor\frac{l+2}{2} \rfloor}\sum_{n=\max(1,3-2q)}^{l+2-2q}\sum_{r,s}\tilde a_{r,s} \frac{|\beta|^{2q}}{n! q!}\frac{1}{2^{q}}\Bigg(\beta^n\sqrt{\frac{k!}{(k-n)!}}F(-k+n,-q;1+n;2)A_{l+2-n-2q}^{k-n}\\+\bar\beta^n\sqrt{\frac{(k+n)!}{k!}}F(-k,-q;1+n;2)A_{l+2-n-2q}^{k+n}\Bigg)+\sum_{n=1}^{l-1}\epsilon_{n}A_{l-n}^k\Bigg)\;, \quad k\neq \nu \ .
\end{multline}
and
\begin{multline}
\epsilon_l\gamma=\sum_{q=2}^{\lfloor\frac{l+2}{2}\rfloor}\sum_{r,s}\tilde a_{r,s} \frac{|\beta|^{2q}}{q!}\frac{1}{2^{q}}F(-\nu,-q;1;2)A_{l+2-2q}^{\nu}
\\+\sum_{q=0}^{\lfloor\frac{l+2}{2} \rfloor}\sum_{n=\max(1,3-2q)}^{l+2-2q}\sum_{r,s}\tilde a_{r,s} \frac{|\beta|^{2q}}{n! q!}\frac{1}{2^{q}}\Bigg(\beta^n\sqrt{\frac{\nu!}{(\nu-n)!}}F(-\nu+n,-q;1+n;2)A_{l+2-n-2q}^{\nu-n}\\+\bar\beta^n\sqrt{\frac{(\nu+n)!}{\nu!}}F(-\nu,-q;1+n;2)A_{l+2-n-2q}^{\nu+n}\Bigg)\;,
\end{multline}
obtained from \eqref{eq:master} by taking $k\neq \nu$ and $k=\nu$ respectively.

The recursion relations can be improved from practical point of view. The appearance of square roots in the formulae slows down significantly the computation when it is implemented in \texttt{Mathematica}, since \texttt{Mathematica} treats irrational parts as if they were unevaluated variables. Fortunately we can eliminate the irrational coefficients simply by rescaling the coefficients $A_l^k$ and by defining
\be
\tilde A_{l}^k=A_{l}^k \sqrt{k!}\;,
\ee
the recurrence equations then become
\begin{multline}
\tilde A_{l}^{k}=\frac{1}{2(k-\nu)}\Bigg(-\sum_{q=2}^{\lfloor\frac{l+2}{2}\rfloor}\sum_{r,s}\tilde a_{r,s} \frac{|\beta|^{2q}}{q!}\frac{1}{2^{q}}F(-k,-q;1;2)\tilde A_{l+2-2q}^{k}
\\-\sum_{q=0}^{\lfloor\frac{l+2}{2} \rfloor}\sum_{n=\max(1,3-2q)}^{l+2-2q}\sum_{r,s}\tilde a_{r,s} \frac{|\beta|^{2q}}{n! q!}\frac{1}{2^{q}}\Bigg(\beta^n{\frac{k!}{(k-n)!}}F(-k+n,-q;1+n;2)\tilde A_{l+2-n-2q}^{k-n}\\+\bar\beta^nF(-k,-q;1+n;2)\tilde A_{l+2-n-2q}^{k+n}\Bigg)+\sum_{n=1}^{l-1}\epsilon_{n}\tilde A_{l-n}^k\Bigg)\;.
\end{multline}
and
\begin{multline}
\gamma\epsilon_l=\frac{1}{\sqrt{\nu!}}\Bigg\{\sum_{q=2}^{\lfloor\frac{l+2}{2}\rfloor}\sum_{r,s}\tilde a_{r,s} \frac{|\beta|^{2q}}{q!}\frac{1}{2^{q}}F(-\nu,-q;1;2)\tilde A_{l+2-2q}^{\nu}
\\+\sum_{q=0}^{\lfloor\frac{l+2}{2} \rfloor}\sum_{n=\max(1,3-2q)}^{l+2-2q}\sum_{r,s}\tilde a_{r,s} \frac{|\beta|^{2q}}{n! q!}\frac{1}{2^{q}}\Bigg(\beta^n \frac{\nu!}{(\nu-n)!}F(-\nu+n,-q;1+n;2)\tilde A_{l+2-n-2q}^{\nu-n}\\+\bar\beta^nF(-\nu,-q;1+n;2)\tilde A_{l+2-n-2q}^{\nu+n}\Bigg)\Bigg\}\;.
\end{multline}
Now by choosing $\gamma=\frac{1}{\sqrt{\nu!}}$ we can get rid of the square roots in the above formula
\begin{multline}
\epsilon_l=\sum_{q=2}^{\lfloor\frac{l+2}{2}\rfloor}\sum_{r,s}\tilde a_{r,s} \frac{|\beta|^{2q}}{q!}\frac{1}{2^{q}}F(-\nu,-q;1;2)\tilde A_{l+2-2q}^{\nu}
\\+\sum_{q=0}^{\lfloor\frac{l+2}{2} \rfloor}\sum_{n=\max(1,3-2q)}^{l+2-2q}\sum_{r,s}\tilde a_{r,s} \frac{|\beta|^{2q}}{n! q!}\frac{1}{2^{q}}\Bigg(\beta^n \frac{\nu!}{(\nu-n)!}F(-\nu+n,-q;1+n;2)\tilde A_{l+2-n-2q}^{\nu-n}\\+\bar\beta^nF(-\nu,-q;1+n;2)\tilde A_{l+2-n-2q}^{\nu+n}\Bigg)\;.
\end{multline}
Notice that this choice sets $\tilde A_0^\nu=\gamma\sqrt{\nu!}=1$, eliminating the irrational factors from the equation.

However, another source of irrational factors can be $\xi$ which appears in the definition of $\beta$. However from \eqref{eq:h_def}, we can see that by defining $\tilde\beta$ to be $\beta=\xi\tilde\beta$ and appropriately rescaling of the coupling, the difference equations can be converted to involve only $\xi^2$ in the imaginary part of $\tilde\beta$. 

A little thought reveals that the difference equation can be setup in such a way that the real part of $\tilde A$-coefficients depend only on the even power of $\xi^2$, while the imaginary part always contains an odd power of $\xi^2$. This means that the real part contains no irrational factors, for a choice of rational choice of all $a_{r,s}$, and, if $\xi^2$ is irrational, the imaginary part of $\tilde A$ will be always proportional to $\xi^2$, multiplying a rational number. 

By splitting the difference equation for $\tilde A$ into its real and imaginary parts, the irrational coefficients appear in a predictable manner, and such treatment of the difference equations whenever the imaginary part of $\tilde A$ is non-vanishing, speeds up the \Mathematica{} algorithm of the package significantly. To activate this feature one needs to call the option \texttt{Imaginary->True} in the \BWDifference{} function, as described in the text.

\section{Proof of uniqueness of minima}\label{sc:uniqueness}

We prove here that the Hamiltonian of the form \eqref{eq:Hamiltonian} with positive $a_{r,s}\ge 0$ has a unique real minimum or no minimum as a function of $x$ and $p$. Equivalently we can also prove that as a function of $X=e^{\sqrt{\hbar}x}$ and $P=e^{\sqrt{\hbar}p}$ there is a unique minimum such that $P>0,X>0$. 

The proof goes as follows. The minimum satisfies the equation
\be
\partial_X\cH=\partial_P\cH=0\;.
\ee
We have that
\be\label{eq:Xmin}
\partial_X\cH=\sum_{r,s}a_{r,s}r X^{r-1}P^s=\sum_{r=r_{min}}^{r_{max}} rX^{r-1}B_r=0
\ee
where
\be
B_r=\sum_s a_{r,s}P^s\;.
\ee
It is clear that for any $P>0$ we have that $B_r\ge0$. Now the equation \eqref{eq:Xmin} can be multiplied by $X^{-r_{min}+1}$ to yield
\be
\sum_{r=r_{min}}^{r_{max}} rX^{r-r_{min}}B_r=0\;.
\ee
Since $B_r>0$, we have two options. If $r_{min}\ge 0$ then the above polynomial has only positive coefficients. If $r_{min}<0$ then the coefficients of $X^k$ with $k<-r_{min}$ have a negative sign, while the rest are positive. We now invoke the rule of Decartes which says that the number $n_p$ of positive real roots of a polynomial is less then or equal to the number of the monomial sign variations of the coefficient. Since this sign variation is 0 or 1, we must have at most one solution.

The same argument can be invoked to show that the equation $\partial_P\cH=0$ has only one solution in $P$ for any $X>0$. This concludes our proof.
	\end{appendix}

	\bibliographystyle{amsmod}
	\bibliography{BenderWu}

\end{document}